\documentstyle[iopl10]{ioplppt}
\input epsf.tex
\def\Re{{\rm Re\ \!}}
\def\Im{{\rm Im\ \!}}
\def\bsigma{\mbox{\boldmath $\sigma$}}
\def\sn{\mathop{\rm sn}\nolimits}
\def\sc{\mathop{\rm sc}\nolimits}
\def\cs{\mathop{\rm cs}\nolimits}
\def\ss{\mathop{\rm ss}\nolimits}
\def\cc{\mathop{\rm cc}\nolimits}
\def\uhp{{\rm uhp}}
\def\pbc{{\rm pbc}}
\def\fbc{{\rm fbc}}
\def\Fbc{{\rm Fbc}} 
\def\Ffbc{{\rm Ffbc}}
\def\FF{{\rm FF}} 
\def\Ff{{\rm Ff}}
\def\simop{\mathop{\sim}}
\def\be{\begin{equation}}
\def\ee{\end{equation}}

\begin{document}
\jl{1}
\title{Bulk and surface properties in the critical phase of
  the two-dimensional $XY$ model}[Properties in the critical phase of
  the two-dimensional $XY$ model]

\author{Bertrand Berche \ftnote {1} {e-mail address: {\tt berche@lpm.u-nancy.fr}} }
\address{Laboratoire de Physique des Mat\'eriaux\ftnote{2}{Unit\'e de
Mixte de Recherche  CNRS No 7556},\\ 
Universit\'e Henri
Poincar\'e, Nancy 1\\
B.P. 239, F-54506 Vand\oe uvre les Nancy, France}
\date{\today}

\begin{abstract}
\baselineskip=9pt
  Monte Carlo simulations of the two-dimensional $XY$ model are performed in
  a square geometry with various boundary conditions (BC). 
  Using conformal mappings 
  we deduce the
  exponent $\eta_\sigma(T)$ of the order parameter correlation
  function and its surface analogue $\eta_\|(T)$  as a function of the temperature 
  in the critical (low-temperature) 
  phase of the model.
  The temperature dependence of both exponents is obtained
  numerically with a good accuracy up to the Kosterlitz-Thouless 
  transition temperature.
  The bulk exponent follows from simulations of correlation functions with
  periodic boundary conditions or order parameter profiles with open
  boundary conditions and with 
  symmetry breaking surface fields.
  At very low temperatures, $\eta_\sigma(T)$  is found in 
  a pretty good agreement with the linear 
  temperature-dependence
  of  Berezinskii's
  spin wave approximation. 
  We also show some evidence that there are no noticeable logarithmic 
  corrections to the behaviour of the
  order parameter density profile at the 
  Kosterlitz-Thouless (KT) transition temperature, while these corrections exist
  for the correlation function.  At the KT transition the 
  value $\eta_\sigma(T_{\rm KT})=1/4$ is
  accurately recovered. The exponent associated to the 
  surface correlations is similarly obtained
  after a slight modification of the boundary conditions: the correlation
  function is computed with free BC, and the profile with
  mixed fixed-free BC.
  It exhibits a monotonous behaviour with
  temperature, starting linearly according to the spin wave approximation and
  increasing up to a value 
  $\eta_\|(T_{\rm KT})\simeq 1/2$ at the Kosterlitz-Thouless
  transition temperature.
  The thermal exponent $\eta_\varepsilon(T)$ is also computed and we give some evidence that
  it keeps a constant value in agreement with the marginality condition of the temperature field
  below the KT transition.
\end{abstract}
\maketitle


\section{Introduction}
\label{sec:intro}
The two-dimensional classical $XY$ model has a rich 
variety of applications in condensed matter physics~\cite{Nelson02}.
It describes of course magnetic films with planar 
anisotropy\ftnote{1}{$2d$ $XY$-type singularities are extremely hard to observe
experimentally in real magnetic layers, since they are washed out by 
crossover effects as soon as a small interlayer interaction takes place in the system.}, 
but also thin-film superfluids or 
superconductors, or two-dimensional solids. 
In statistical physics, this model was also extensively studied
for fundamental reasons, as describing for example classical Coulomb gas or
fluctuating surfaces and the roughness transition. 
Although no exact solution exists for this model,
many of its essential properties are known from different approaches.

The model undergoes a standard temperature-driven 
paramagnetic to ferromagnetic phase 
transition in $d>2$, characterised e.g. by a 
power-law divergence of the correlation length near criticality,
\be
\xi\sim|t|^{-\nu},
\ee
(see e.g. Ref.~\cite{HasenbuschToeroek99}).
In two dimensions, it exhibits a rather different behaviour, 
since long-range order of spins with continuous
symmetries (if there is no source of anisotropy in spin space) 
is forbidden according to  the 
Mermin-Wagner theorem~\cite{MerminWagner66,Hohenberg67}. 
Indeed there is no  spontaneous magnetisation at finite 
temperature in two-dimensional
systems having a continuous symmetry group, for example
$XY$ or $O(2)$ spin models, since order would 
otherwise be destroyed by spin wave
excitations~\cite{Berezinskii71}. On the other hand, there is an infinite 
susceptibility at low temperatures,  
indicating another type of transition~\cite{StanleyKaplan66}.
Such systems can indeed display a low-temperature phase 
with quasi-long-range order and  
a defect-mediated transition towards a usual 
paramagnetic phase at high temperatures. This transition is governed by unbinding 
of topological defects.

We now specify the case of the $2d$ $XY$-model: 
consider a square lattice with two-components spin variables 
$\bsigma_w=(\cos\theta_w,\sin\theta_w)$, $|\bsigma_w|^2=1$, 
located at the sites 
$w$ of a lattice $\Lambda$ of linear extent $L$, 
and interacting through the usual nearest-neighbour ferromagnetic interaction
\begin{eqnarray}
        -\frac{H}{k_BT}&=&K\sum_{w}\sum_\mu\bsigma_w\cdot\bsigma_{w+\hat\mu}
        \nonumber\\
        &=&K\sum_{w}\sum_\mu\cos(\theta_w-\theta_{w+\hat\mu}),
        \label{Ham}
\end{eqnarray}
where $K=J/k_BT$, with $J$ the ferromagnetic coupling strength and
$\mu=1,2$ labels the directions in $\Lambda$ ($\hat\mu$ is the corresponding unit vector).
The angle $\theta_w$ measures the deviation of $\bsigma_w$ from an arbitarty reference direction.

The low-temperature regime, 
governed by spin-wave excitations, 
was investigated  
by Berezinskii~\cite{Berezinskii71} in the harmonic 
approximation. 
In the low-temperature limit (also called spin-wave phase), 
localised non-linear excitations (called vortices)
are indeed bounded in pairs of zero vorticity, 
and thus do not affect crucially 
the spin-wave description. At a first approximation
the effect of vortices can be completely neglected and the spin wave 
approximation, after expanding the cosine,  reproduces the main features of the 
low-temperature physics. This harmonic approximation is justified, provided 
that the spin disorientation remains small, i.e.
at sufficiently low temperature:
\be
-\frac{H}{k_BT}\simeq 
-\frac{H_0}{k_BT}-{\textstyle\frac 12}K\sum_{w}\sum_\mu(\theta_w-\theta_{w+\hat\mu})^2.
\ee
Within this approximation, the quadratic 
energy corresponds to a Gaussian equilibrium 
distribution and the two-point correlation function at inverse temperature
$\beta$ becomes~\cite{Rice65,Wegner67,Sarma72,Tsallis76}
\begin{eqnarray}\fl
&Z_\beta=&\left(\prod_{w}\int\frac{d\theta_w}{2\pi}\right)\prod_{w,\mu}
\exp[-{\textstyle\frac 12}K(\theta_w-\theta_{w+\hat\mu})^2],\\
\langle\bsigma_{w_1}\cdot\bsigma_{w_2}\rangle&\simeq&
Z_\beta^{-1}\prod_w\int\frac{d\theta_w}{2\pi}
\cos(\theta_{w_1}-\theta_{w_2})\times {\rm e}^{-\frac 12K
\sum_{w}\sum_\mu(\theta_w-\theta_{w+\hat\mu})^2}\nonumber \\
&\simeq&
{\rm e}^{
-{\textstyle\frac 12}
Z_\beta^{-1}\prod_w\int\frac{d\theta_w}{2\pi}
(\theta_{w_1}-\theta_{w_2})^2\times {\rm e}^{-\frac 12K
\sum_{w}\sum_\mu(\theta_w-\theta_{w+\hat\mu})^2}
}
\nonumber \\
&\simeq&|w_1-w_2|^{-1/2\pi K},
\end{eqnarray}
hence 
\be\eta_\sigma^{SW}(T)=\frac{1}{2\pi K}=\frac{k_BT}{2\pi J}.\label{eta-SW}\ee 

When the temperature increases, the bounded vortices  appear in increasing 
number. Their influence becomes more prominent, producing
a deviation from the linear spin-wave 
contribution in equation~(\ref{eta-SW}), but the  
order parameter correlation function still
decays algebraically with an exponent $\eta_\sigma(T)$ which depends on the
temperature.  
The transition eventually takes place 
at a temperature $T_{\rm KT}$ 
(usually called Berezinskii-Kosterlitz-Thouless critical
temperature) when the pairs dissociate, thereby
destroying short-range order in the system.
The mechanism of unbinding of vortices was studied by
Kosterlitz and Thouless~\cite{KosterlitzThouless73,Kosterlitz74,Villain75} 
using approximate
renormalization group methods. 
Above $T_{\rm KT}$ (in the vortex phase), 
the correlation function 
recovers a usual exponential decay.
This very peculiar topological transition is characterised by 
essential singularities 
when approaching the critical point from the high temperature phase, $t=T-T_{\rm KT}\to 0^+$,
\begin{eqnarray}
  \xi&\sim&{\rm e}^{b_\xi t^{-\sigma}},\label{eq-nu}\\
  \chi&\sim&{\rm e}^{b_\chi t^{-\sigma}}\sim\xi^{2-\eta_\sigma},
\label{eq-eta}
\end{eqnarray}
whence $\eta_\sigma\equiv \eta_\sigma(T_{\rm KT})=2-b_\chi/b_\xi$. 
For reviews, see e.g. 
Refs.~\cite{Nelson02,KosterlitzThouless78,ItzyksonDrouffe89,Cardy96,GulasciGulasci98}. 

The existence of a scale-invariant power-law decay of the correlation function
implies that at each value of the temperature 
there corresponds a fixed point, or equivalently
that there is a continuous 
line of fixed points at low temperatures. The temperature is thus 
a marginal field in the spin-wave phase which becomes relevant in the 
vortex phase.

Not much is known in the intermediate regime
between the spin wave approximation
at low temperature and the Kosterlitz-Thouless 
results at the topological transition
and in fact the precise determination of the critical behaviour 
of the two-dimensional 
$XY$ model in the critical
phase remains a challenging problem. Most of the studies were dedicated to the
determination of critical properties at $T_{\rm KT}$\ftnote{2}{When approaching
the transition from above, the value of $T_{\rm KT}$ is often obtained by 
fixing $\sigma=1/2$ in equation (\ref{eq-nu}), and $\eta_\sigma(T_{\rm KT})$ then 
follows from a fit
of the susceptibility to equation (\ref{eq-eta}).}.
Many results were obtained using 
Monte Carlo simulations (see e.g. 
Refs.~\cite{FernandezEtal86,GuptaEtAl88,WeberMinnhagen88,BifferalePetronzio89,Wolff89,GuptaBaillie92,JankeNather93,Olsson95,KennaIrving95,Kim96,Janke97,KennaIrving97})
at $T_{\rm KT}$ and slightly above, or 
high-temperature series expansions~\cite{ButeraComiMarchesini89,ButeraComi94},
but the analysis was made difficult by the existence of logarithmic 
corrections, e.g.
\be
\chi\sim\xi^{2-\eta_\sigma}(\ln \xi)^{2\theta}\left[1+O\left(\frac{\ln\ln\xi}{\ln\xi}\right)\right]
\ee
in the high-temperature
regime, or 
\be\langle\bsigma_{w_1}\cdot\bsigma_{w_2}\rangle
\simeq \frac{(\ln |w_1-w_2|)^{2\theta}}{|w_1-w_2|^{\eta_\sigma}}
\left[1+O\left(\frac{\ln\ln |w_1-w_2|}{\ln|w_1-w_2|}\right)\right].
\label{eq9}\ee
 at $T_{\rm KT}$ exactly~\cite{ItzyksonDrouffe89,AmitGoldschmidtGrinstein80,ZisookKadanoff80},
with a probable value of $\theta=\frac{1}{16}$~\cite{CampostriniPelissettoRossiVicari96}. 
Due to these logarithmic corrections which make the fits quite difficult to achieve safely, 
the values of $T_{\rm KT}$ and 
$\eta_\sigma(T_{\rm KT})$ were a bit controversial as shown in table~1
of reference~\cite{KennaIrving97}.
The resort to large-scale simulations was then 
needed in order to confirm this picture~\cite{Janke97}.
Although essentially dedicated to the KT point, some of these papers 
also report the value of $\eta_\sigma(T)$ computed from the correlation
function decay  at several temperatures
in the critical 
phase~\cite{FernandezEtal86,BifferalePetronzio89,Wolff89,GuptaBaillie92},  
while a systematic study of the helicity modulus at low temperature 
previously led more directly to the same 
information~ \cite{Himbergen84}.

Recently, we proposed a rather different 
approach~\cite{BercheFarinasParedes02,Berche02} which only 
requires a moderate computation effort. 
Assuming that the low-temperature phase exhibits all the 
characteristics of a critical
phase with conformal invariance~\cite{Henkel99}
 (invariance under rotation, translation and scale transformations,
short-range interactions, isotropic scaling), 
we use the covariance law of $n-$point correlation
functions under the mapping of a two-dimensional system confined inside a 
square onto the infinite or half-infinite
plane. 
The scaling dimensions are then obtained through a simple 
power-law fit where the shape
effects are encoded in the conformal mapping, via the definition of a rescaled distance variable
adapted to the description of the confined geometry.

This technique is well known
since the appearance of conformal invariance. Conformal mappings have been extensively 
used, mainly in the case of Ising or Potts models, in order to investigate the 
critical properties in restricted geometries.
Density profiles or correlations  have been investigated in
various confined systems (surfaces~\cite{Cardy84,ItzyksonDrouffeIX89}, 
corners~\cite{BarberPeschelPearce84,DaviesPeschel91}, 
strips~\cite{BurkhardtEisenriegler86,BurkhardtXue91a,BurkhardtXue91b,TurbanIgloi97},
squares~\cite{BurkhardtDerrida85,KlebanEtal86} 
or parabolic 
shapes~\cite{PeschelTurbanIgloi91,BlawidPeschel94,BercheDebierreEckle94}
for a review,
see~\cite{IgloiPeschelTurban93}). 
The exact expression of the two-point correlation functions have also been 
calculated on the torus  for the pure 
Ising case~\cite{diFrancescoSaleurZuber87}, and large scale
Monte Carlo simulations of the $2d$ Ising model have been shown to
reproduce the correct behaviour, including corrections due to
the finite size of the 
lattice~\cite{TalapovAndreichenkoDotsenkoShchur93,TalapovShchur94}.

In this paper, we report a systematic application of conformal
mappings to the investigation of critical properties of the $2d$ $XY$ model
in the spin-wave phase. The physical quantities are obtained using Monte Carlo
simulations in square shaped confined systems.
We may then determine accurately the correlation function exponent 
$\eta_\sigma(T)$ in the 
low-temperature phase $T\le T_{\rm KT}$~\cite{BercheFarinasParedes02} 
but also, by choosing convenient boundary conditions (BC),
the surface critical exponent which describes the decay of the 
correlation function parallel to a free surface,
$\eta_\parallel(T)$~\cite{Berche02}. The thermal exponent $\eta_\varepsilon$ will also
be considered.
Section~\ref{sec:method} presents the expected functional expressions of
correlation functions and density profiles with various BC, and use of these
expressions is systematically reported in section~\ref{sec:sim}. It is worth noticing already
that the results following the analysis of the two-point correlation functions are not fully convincing,
but the main conclusions are then supported by the most refined investigations of the density profiles.
A discussion of the results is given in the last section.

\section{Functional expressions of correlation functions and density profiles in
a square-shaped critical system}
\label{sec:method}
\subsection{Preliminary and notations}
In the following, we shall specify two different types of geometries, 
depending on our purpose. They are related through a conformal
transformation. The critical properties in the first system will be 
reached through the study of the confined system in the second one.
\begin{description}
\item[i)] The infinite plane 
\begin{eqnarray}
z=x+iy,\nonumber\\
  -\infty< \Re z < +\infty,\nonumber\\ 
  -\infty<\Im z <+\infty
\end{eqnarray} 
or the upper half-plane 
\begin{eqnarray}
z=x+iy,  \nonumber\\
  -\infty< \Re z < +\infty,\nonumber\\
 y=\Im z\ge 0.
\end{eqnarray} 
This is the system which
  is eventually interesting in the thermodynamic limit. It is of course
  not directly accessible from numerical techniques.
  In the second case, the existence of a free surface materialised
  here by the real axis will allow to define both bulk and surface quantities.
\item[ii)] The square 
\begin{eqnarray}
w=u+iv,  \nonumber\\
 -L/2\le \Re w\le L/2,\nonumber\\
0\le \Im w \le L.
\end{eqnarray} 
  This is the natural geometry
  for Monte Carlo simulations. The boundary conditions may be open
  at the edges (square) or periodic (torus). In the following, we will obtain 
  numerical results in the square or the torus, and our
  aim will be to translate them into their half-infinite or infinite 
  system counterpart\ftnote{3}{Here and in the following, $z$ will 
constantly refer to a complex coordinate in the plane, while $w$ will refer to the
complex coordinate in the square geometry. In the plane, \uhp\  specifies
the upper half-plane.  In the square, \pbc, \fbc, \Fbc, and \Ffbc\ means that the
boundary conditions are respectively periodic, free, fixed, or mixed 
fixed-free according to a description which will be given later. }.
\end{description}

It is worth referring also to
 {the infinitely long strip} of width $L$, $k+il$, 
  $-\infty <k< +\infty$, $0\le l\le L$: this geometry is incidentally
  mentioned here because the choice of the boundary conditions is more
  obvious there than in the square system. 
  Although no computation will be performed in this
  geometry, we will refer to it when introducing the profiles. 
  It is particularly interesting, since many results 
  concern such strips, e.g. when obtained with transfer
  matrices~\cite{BloteNienhuis89} or quantum chains~\cite{AlltonHamer88} 
  which correspond to the Hamiltonian 
  limit of classical systems confined
  in strips.   In the transverse direction ($l-$coordinate), the boundary 
  conditions  may be open
  (strip) or periodic (cylinder).
  A very accurate numerical study of the $O(n)$ model (including
  $XY$) using the transfer matrix of a related loop gas model 
  was for example reported by Bl\"ote and 
  Nienhuis~\cite{BloteNienhuis89}. The periodic strip is obtained from the 
  plane through the standard logarithmic mapping $k+il=\frac{L}{2\pi}\ln z$
  which implies an asymptotic exponential decay for the correlation functions
  along the strip at criticality, 
  \be\langle\bsigma_{k_1+il}\bsigma_{k_2+il}\rangle_{\pbc}
  \sim\exp (-2\pi x_\sigma |k_1-k_2|/L).\label{eq-corrstrip}\ee
  The scaling dimensions are thus determined by the universal correlation 
  length amplitudes, $\xi_\sigma^{-1}=2\pi x_\sigma /L$, which are given, in 
  the transfer matrix formalism, by the two leading eigenvalues of the transfer matrix 
  $\xi_\sigma^{-1}=\ln(\Lambda_0/\Lambda_1)$. Bl\"ote and Nienhuis obtained
  the most precise values of the thermal and magnetic scaling dimensions
  at the KT transition, $x_\varepsilon=\frac 12\eta_\varepsilon=
  2.000~000~(2)$ and $x_\sigma=\frac 12\eta_\sigma=
  0.125~000~(1)$.
In the following, we will exploit expressions similar to equation~(\ref{eq-corrstrip}), transposed
in the square geometry. We mention here that our approach proceeds from Monte Carlo 
simulations and cannot pretend to an accuracy comparable to the results of
Bl\"ote and Nienhuis.

\subsection{Conformal rescaling of the correlation functions}
In an infinite system at criticality, the correlation function 
is known to behave asymptotically as
\be
\langle \bsigma_{z_1}\cdot\bsigma_{z_2}\rangle\sim |z_1-z_2|^{-\eta_\sigma}.
\label{eq0}\ee
This defines the bulk correlation function exponent $\eta_\sigma$.
The surface critical properties are also interesting and extra universal critical exponents can be measured
from simulations performed in finite systems with open boundary conditions.
In the semi-infinite geometry $z=x+iy$ 
(the free surface being defined by the $x$ axis),  
the two-point correlation function is fixed up
to an unknown scaling function.
Fixing one point $z_1$ close to the free surface 
($z_1=i$) of
the upper half-plane (\uhp), 
and leaving the second point $z_2$ explore the
rest of the geometry, the following behaviour is expected:
\be
\langle \bsigma_{z_1}\cdot\bsigma_{z_2}\rangle_{\uhp}
\sim (y_1y_2)^{-x_\sigma}\psi(\omega),\label{eq:G}
\ee
with $\eta_\sigma=2x_\sigma$. The dependence on 
\be\omega=\frac{y_1y_2}{\mid z_1-z_2\mid^2}\label{eq-omega}\ee 
of the universal scaling function $\psi$ is constrained by the special conformal transformation~\cite{Cardy84,ItzyksonDrouffeIX89}, and its asymptotic 
behaviour is implied by scaling e.g.  
$\psi(\omega)\sim\omega^{x_\sigma^1}$ when 
$y_2\gg 1$, with $x_\sigma^1=
{\textstyle\frac 12}\eta_\parallel$
the magnetic surface scaling dimension. 

In a square system, the order parameter correlation function 
$\langle\bsigma_{w_1}\cdot\bsigma_{w_2}\rangle$
is strongly affected by boundary and shape effects and thus deviates
significantly from the thermodynamic limit expressions~(\ref{eq0}) and
(\ref{eq:G}).
It should obey a scaling
form which reproduces the expected power-law behaviours, for example 
\begin{equation}
 \langle\bsigma_{w_1}\cdot\bsigma_{w_2}\rangle =|w_1-w_2|^{-\eta_\sigma}
  f_{\rm sq.}(w_1/L,w_2/L),
  \label{eq:scalG}
\end{equation}
where the function $f_{\rm sq.}$ encodes shape effects in a very
complicated manner which also depends on the boundary conditions, 
so that the connection between equations~(\ref{eq0}) or (\ref{eq:G}) and (\ref{eq:scalG}) is not
obvious. 
Conformal invariance provides an efficient technique to avoid these shape
effects, or at least, enable to include explicitly the shape
dependence in the functional expression of the correlators through the 
conformal
covariance transformation under a mapping $w(z)$:  
\be
\langle\bsigma_{w_1}\cdot\bsigma_{w_2}\rangle=|w'(z_1)|^{-x_\sigma}|w'(z_2)|^{-x_\sigma}
\langle\bsigma_{z_1}\cdot\bsigma_{z_2}\rangle.
\label{covconf}
\ee
Conformal invariance holds in isotropic systems with short
range interactions which exhibit translation, rotation, as well as scale
invariance. Usually these conditions are so restrictive that they can only
be valid at the critical point, and we assume here that 
the general covariance equation~(\ref{covconf}) holds
in the whole $T<T_{\rm KT}$ phase.

In order to get a functional expression of the correlation function
inside the square geometry $w$, one simply has to use the convenient
mapping $w(z)$:
\begin{description}
\item[i)]{\bf Conformal mapping of the plane onto the torus:}
in the continuum limit, the correlation function of the Ising model
on the torus is exactly 
known~\cite{diFrancescoSaleurZuber87,ItzyksonDrouffeIX89,diFrancescoMathieuSenechal97}
\be\fl
\langle\bsigma_{w_1}\cdot\bsigma_{w_2}\rangle_{\pbc}
\simop_{\rm Ising}\left[\sum_{\nu=2}^4\left|\theta_\nu(0)\right|\right]^{-1}
\sum_{\nu=1}^4\left|\theta_\nu\left(\frac{w_{12}}{2L}\right)\right|\times
\left|\theta'_1(0)/\theta_1\left(\frac{w_{12}}{L}\right)
\right|^{1/4},
\label{eq-corrIsingtorus}\ee
($\eta_\sigma=1/4$) 
where $\theta_\nu(\alpha)$ are the Jacobi theta 
functions and $w_{12}=w_1-w_2$. 
A similar expression exists for the
energy-energy correlations~\cite{ItzyksonDrouffeIX89}. 
It can be generalised to a critical system
characterised by any correlation function exponent $\eta_\sigma$. For
simplicity, we fix in the following one
point ($w_1$) as the origin, and the second point is simply written $w$.
The dependence on the relative distance $w$ in the above expression,
\be
\langle\bsigma_{0}\cdot\bsigma_{w}\rangle_{\pbc}
\sim 
\sum_{\nu=1}^4\left|\theta_\nu\left({w}/{2L}\right)\right|\times
\left|\theta_1\left({w}/{L}\right)
\right|^{-1/4},\ee 
is compatible with the general covariance 
equation~(\ref{covconf}), rewritten as
\be\langle\bsigma_{0}\cdot\bsigma_{w}\rangle_{\pbc}\sim|w'(z)|^{-\eta_\sigma/2}
[z(w)]^{-\eta_\sigma},\ee 
provided that $|z(0)|\ll |z(w)|$ (large distance behaviour). 
The r.h.s. is a function of $w$ hereafter
denoted by
$[\zeta(w)]^{-\eta_\sigma}$, with $\eta_\sigma=1/4$ and the rescaled distance is
\be
\zeta(w)=\left|\theta_1\left({w}/{L}\right)
\right|\times\left(\sum_{\nu=1}^4\left|\theta_\nu\left({w}/{2L}\right)
\right|\right)^{-4}.
\label{eq-varpi}
\ee
Therefore equation~(\ref{eq-corrIsingtorus}) can be generalised to any 
correlation function of a critical system
inside a torus,
\be
\langle\bsigma_{0}\cdot\bsigma_{w}\rangle_{\pbc}
\sim [\zeta(w)]^{-\eta_\sigma}.
\label{eq-corr-torus}
\ee
\begin{figure} [th]
\vspace{0.2cm}
        \epsfxsize=12cm
        \begin{center}
        \mbox{\epsfbox{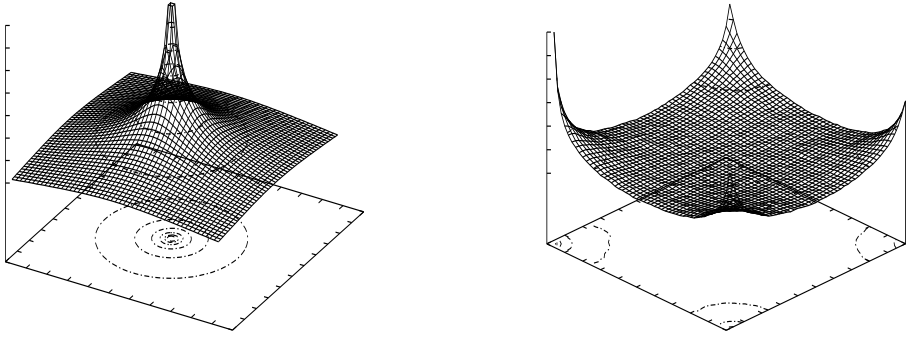}\qquad}
        \end{center}
        \caption{Correlation functions in the plane (left) and in the square
        ($\pbc$)  geometries (right).}
        \label{fig_correlTC}  \vskip -0cm
\end{figure}

A sketch of the correlation functions in the plane and in the square
geometries is shown in figure~\ref{fig_correlTC}.

\item[ii)]{\bf Conformal rescaling of the upper half-plane inside a square:}
the conformal transformation of the half-plane $z=x+iy$ ($0\le y<\infty$) 
inside a square $w=u+iv$ of size
$L\times L$ ($-L/2\le u\le L/2$, $0\le v\le L$)
with open boundary conditions
along the four edges is realized by a Schwarz-Christoffel transformation~\cite{LavrentievChabat}
\be
w(z)={L\over 2{\rm K}}{\rm F}(z,k),
\quad z={\rm sn}\left({2{\rm K}w\over L}\right).
\label{eq-SchChr}
\ee
Here, $F(z,k)$ is the elliptic integral of the 
first kind, 
\be
F(z,k)=\int_0^z[(1-t^2)(1-k^2t^2)]^{-1/2}{\rm d}t,
\ee
${\rm sn}\ \! (2{\rm K}w/ L)$ 
the Jacobian elliptic sine, ${\rm K}=K(k)=F(1,k)$ the
complete elliptic integral of the first kind, 
and the modulus $k$ depends on the aspect ratio
of $\Lambda$  and is here solution
of $K(k)/K(\sqrt{1-k^2})=\frac 12$:
\begin{equation}
  k=4\left(\frac{\sum_{p=0}^\infty q^{(p+1/2)^2}}{1+2
  \sum_{p=1}^\infty q^{p^2}}\right)^2\simeq 0.171573 ,\quad q={\rm e}^{-2\pi}.
  \label{eq: k}
\end{equation}
Using the mapping~(\ref{eq-SchChr}), one obtains the local rescaling factor in 
equation~(\ref{covconf}),
$w'(z)=\frac{L}{2{\rm K}}[(1-z^2)(1-k^2z^2)]^{-1/2}$,
and inside the square, keeping $w_1\sim O(1)$ fixed,  
the two-point correlation function 
becomes (see e.g. Ref.~\cite{ChatelainBerche99})
\begin{eqnarray}  
&&\langle\bsigma_{0}\cdot\bsigma_w\rangle_{\fbc}
\sim[\kappa(w)]^{-\frac 12\eta_\sigma} \psi(\omega)\nonumber\\
  &&      \kappa(w)= {\rm Im}\left[{\rm sn}\frac{2{\rm K}w}{L} \right]\times
        \left| \left(1-{\rm sn}^2\frac{2{\rm K}w}{L}
        \right)  \left( 1-k^2{\rm sn}^2\frac{2{\rm K}w}{L} \right)\right|^{-1/2},
\label{eqkappa}
\end{eqnarray} 
where $\fbc$ specifies that the open square has free boundary conditions.
This expression is correct up to a constant amplitude determined by
$\kappa(w_1)$ which is kept fixed, but the function $\psi(\omega)$ is still
varying with the location of the second point, $w$. 

A sketch of the correlation functions in the plane and in the square
geometries is shown in figure~\ref{fig_correlTCdemiplan}.
\begin{figure} [ht]
\vspace{0.2cm}
        \epsfxsize=12cm
        \begin{center}
        \mbox{\epsfbox{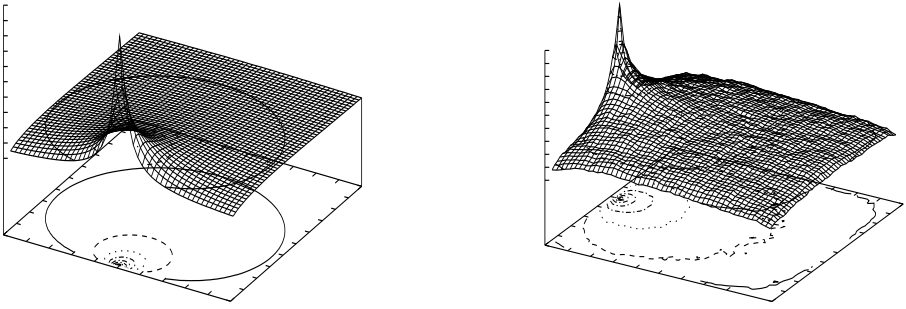}\qquad}
        \end{center}
        \caption{Correlation functions in the upper half-plane (left) 
          and in the square
        ($\fbc$)  geometries (right).}
        \label{fig_correlTCdemiplan}  \vskip -0cm
\end{figure}

\end{description}

\subsection{Density profiles in restricted geometries}
\begin{figure} [ht]
\vspace{0.2cm}
        \epsfxsize=12cm
        \begin{center}
        \mbox{\epsfbox{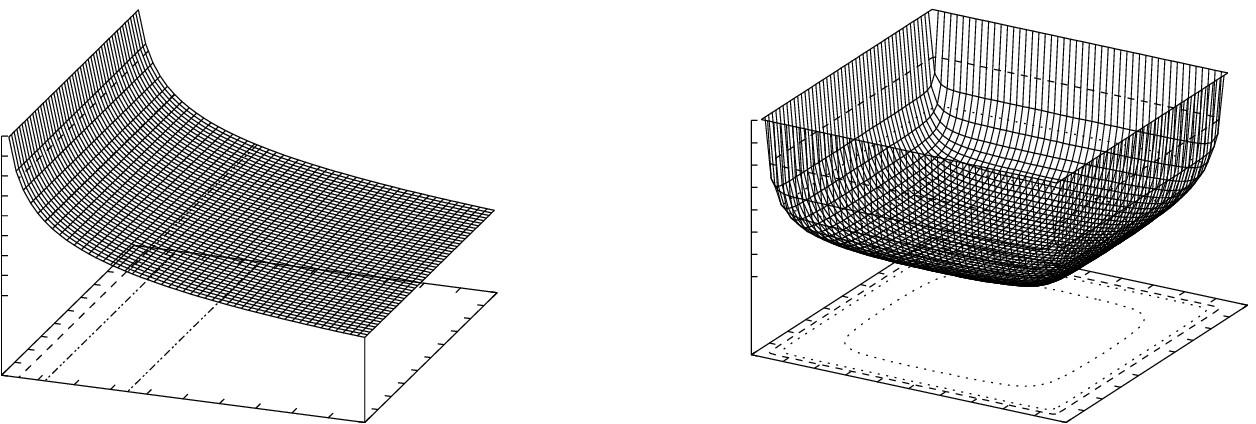}\qquad}
        \ \vspace*{10mm}\ 
        \epsfxsize=12cm
        \mbox{\epsfbox{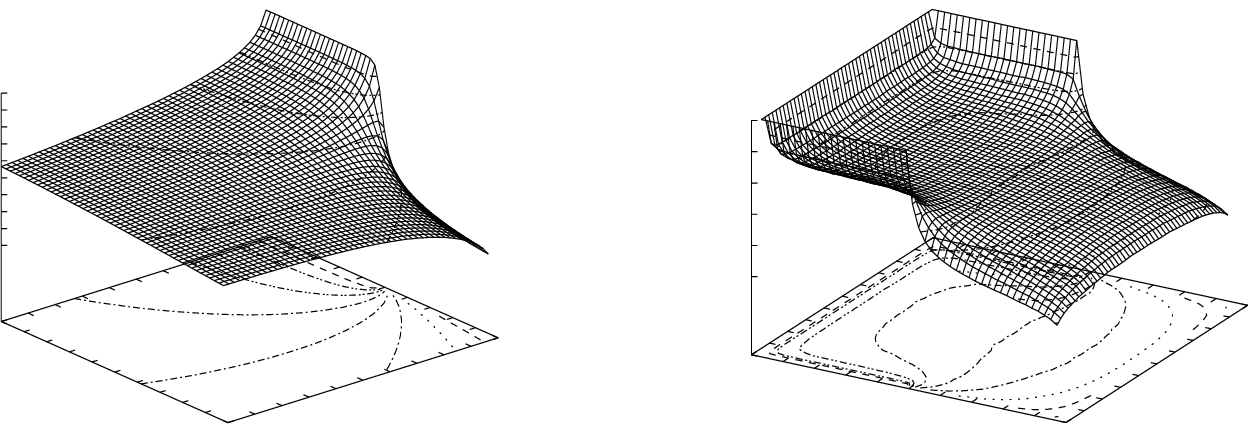}\qquad}
        \end{center}
        \caption{Order parameter profile in the plane (left) and in the square
        geometries (right) with fixed boundary conditions (top) and 
        mixed fixed-free BC (bottom).}
        \label{fig_profilTCFxd}  \vskip -0cm
\end{figure}

In order to cancel the role of the unknown scaling function in the half-plane
geometry in equation~(\ref{eqkappa}),
it is more convenient to work with a density profile $m(w)$ in the presence
of symmetry breaking surface fields ${\bf h}_{\partial\Lambda}$ 
on the boundary $\partial\Lambda$ of the lattice 
$\Lambda$. This is a
one-point correlator which
scales
in the half-infinite geometry as:
\be 
m(z)=\langle\bsigma_z\cdot{\bf h}_{\partial\Lambda(z)}\rangle_{\uhp} 
= {\rm const\times} y^{-x_\sigma}\label{eq21}\ee
and it maps onto (see figure~\ref{fig_profilTCFxd})
\be
m_{\Fbc}(w)=\langle\bsigma_w\cdot{\bf h}_{\partial\Lambda(w)}
\rangle_{\Fbc} =
        {\rm const\times}[\kappa(w)]^{-\frac 12\eta_\sigma}
\label{eqParamOrder}        \ee
where the function $\kappa(w)$ defined in equation~(\ref{eqkappa}) again 
comes from the mapping and $\Fbc$ means that the open square has fixed
boundary conditions (figure~\ref{fig_profilTCFxd}).

Equation~(\ref{eq21}) is a special case of conformally invariant 
profiles predicted e.g. by
Burkhardt and Xue~\cite{BurkhardtXue91a,BurkhardtXue91b} in strip geometries 
$k+il$ or in their half-infinite counterpart $z=x+iy$:
\be m_{ab}(z)=y^{-x_\sigma}F_{ab}(\cos\theta).\ee
The scaling function $F_{ab}(x)$ depends on the 
universality class and on the boundary conditions
denoted by $a$ and $b$~\cite{FisherdeGennes78}. This notation is reminiscent
from the strip geometry where $a$ and $b$ specifies the two boundaries
of the strip at $l=0$ and $l=L$ respectively.
In the case of fixed BC  (denoted by $\Fbc$ in the square and $\FF$ in the strip) 
corresponding to 
equation~(\ref{eq21}) and (\ref{eqParamOrder}), one gets 
\be
m_{\Fbc}(z)=y^{-\eta_\sigma/2},
\ee
while in the case of fixed-free BC  ($\Ffbc$ in the square or more naturally
$\Ff$ in the strip)\ftnote{4}{The spins located on the $x>0$ half axis ($l=0$ strip surface) are kept 
fixed while those of the $x<0$ half axis ($l=L$ strip surface) are free.} 
Burkhardt and Xue have shown in the case of Ising and Potts
models that
\be
m_{\Ffbc}(z)=y^{-\eta_\sigma/2}(\cos{\scriptstyle\frac {1}{2}}
\theta)^{\eta_\parallel/2}.
\ee
This expression was since then  mainly used in the strip 
geometry~\cite{refconf,CarlonIgloi98}.
In the square geometry, the fixed-free BC correspond to keeping the spins fixed for $u>0$, and
the profile is expected to obey the following 
ansatz~\cite{Berche02} (figure~\ref{fig_profilTCFxd}):
\begin{eqnarray}  
&&m_{\Ffbc}(w)\sim {\rm const}\times 
[\kappa(w)]^{-\frac 12\eta_\sigma}[\mu(w)]^{\frac 12\eta_\|}, \nonumber\\
 &&\mu(w)= \frac {1}{\sqrt 2}\left(1+{\rm Re}\left[{\rm sn}\frac{2{\rm K}w}{L} \right]\times
        \left|\  \!{\rm sn}\frac{2{\rm K}w}{L}\right|^{-1}\right)^{1/2}
\label{eqkappamu}
\end{eqnarray}

In this paper we will essentially apply equations~(\ref{eq-corr-torus}),
(\ref{eqkappa}), (\ref{eqParamOrder}) and (\ref{eqkappamu}) in order
to determine $\eta_\sigma$ and $\eta_\parallel$.

\section{Monte Carlo simulations}
\label{sec:sim}
\subsection{Description of the algorithm}
Simulations of $2d$ $XY$-spins are performed using Wolff's cluster
Monte Carlo algorithm~\cite{Wolff89a}. 
The boundary conditions may be periodic, free or fixed. When the boundaries
are left periodic or free, there is no particular problem, but 
a relatively
precise numerical determination of the order parameter profile of the 
$2d$ $XY$ model
confined inside a square
with fixed boundary conditions (playing the r\^ole of
ordering surface fields ${\bf h}_{\partial\Lambda(w)}$) introduces {\it a priori} 
a technical difficulty which is easily circumvented. In practice, 
the symmetry is broken by keeping the  boundary spins located 
along the four edges of the square fixed, 
e.g. $\bsigma_w=(1,0), \forall w\in\partial\Lambda$ during the Monte Carlo 
simulation. The Wolff algorithm should thus
become less efficient, since close to criticality the unique cluster will often
reach the boundary and no update would be made in this case. To prevent this, 
we use the symmetry
of the Hamiltonian~(\ref{Ham}) under a global rotation of all the spins. Even when the cluster 
reaches the fixed boundaries
$\partial\Lambda(w)$, 
it is updated, and the order parameter profile is then measured with 
respect to the common
new direction  of the boundary
spins, $m_{\Fbc}(w)=\langle\bsigma_w\cdot\bsigma_{\partial\Lambda(w)}\rangle_{\Fbc}$.
The new configuration reached would thus 
correspond - after a global rotation of all the spins 
of the system to re-align
the boundary spins in their original $(1,0)$ direction - to a
a new configuration of equal total energy
and thus the same statistical weight as the one actually produced.

In the literature, many papers were dedicated to the study of the 
characteristic properties of the Kosterlitz-Thouless transition. The value
of $T_{\rm KT}$ will be considered as known with a good 
accuracy~\ftnote{5}{The numerical value of the transition temperature has been
deduced from different approaches, e.g. high-temperature series expansions (e.g. in \cite{ButeraComi94}),
temperature dependence of the correlation length (e.g. in \cite{GuptaBaillie92}), zeroes of the partition 
function (e.g. in \cite{KennaIrving97}), jump in the helicity modulus (e.g. in \cite{WeberMinnhagen88}),
\dots and the recent estimations are compatible
with each other.}. We will take here the value reported by Gupta and 
Baillie~\cite{GuptaBaillie92},  or Campostrini et al~\cite{CampostriniPelissettoRossiVicari96}, 
$k_BT_{\rm KT}/J=0.893(1)$.

\subsection{Behaviour of the correlation functions}
\subsubsection{Scaling of correlations on the torus:}
Simulations are performed on a square with periodic boundary conditions.
Here we choose systems of size $L=100$. After thermalization
($10^6$ iterations) and computation ($10^6$ other iterations), we get the
spin orientations $\{\theta_w\}$  at all the lattice sites.
The correlation function is defined as  
\be
\langle\bsigma_0\cdot\bsigma_w\rangle_{\pbc}=\langle\cos(\theta_0-\theta_w)
\rangle_{\pbc},
\ee
where $0$ is an arbitraty reference site (e.g. a corner in figure~\ref{fig_correlTC}).

\begin{figure} [ht]
\vspace{0.2cm}
        \epsfxsize=10cm
        \begin{center}
        \mbox{\epsfbox{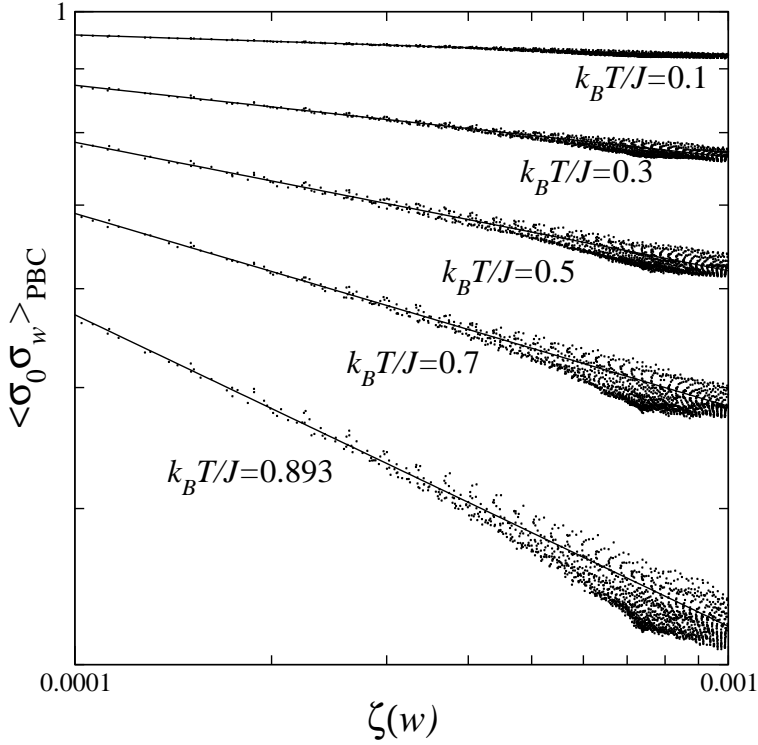}\qquad}
        \end{center}
        \caption{Log-log plot of the order parameter correlation function
        in a square system with periodic boundary conditions
        {\it vs} the rescaled variable $\zeta(w)$. The data correspond
        to a quarter of the lattice $\Lambda$, for $0<u\le L/2$ and
        $0<v\le L/2$, $L=100$. Four different 
        temperatures in the spin-wave phase are shown ($k_BT/J=0.1$, 0.3, 0.5, and
        0.7 from top to bottom), and at the KT transition temperature.}
        \label{fig-corrPbc}  \vskip -0cm
\end{figure}

It has to be fitted to equation~(\ref{eq-corr-torus}) with
$\zeta(w)$ given by equation~(\ref{eq-varpi}). Expansions of the 
$\theta$-functions can be found in the literature,
e.g. in Abramowitz and Stegun~\cite{AbramowitzStegun70},
\be
  \theta_1(w/L)=2{\rm e}^{-\pi/4}\sum_{n=0}^\infty
  (-1)^n{\rm e}^{-n(n+1)\pi}\sin\left[\frac{(2n+1)\pi w}{L}\right]
\ee
and similar expression for $\theta_2$, $\theta_3$ and $\theta_4$. Here it has been taken
into account the fact that the system is inside a square (not a rectangle $L\times L'$), therefore the 
expansion parameter  ${\rm e}^{-\pi L'/L}$ reduces to $q={\rm e}^{-\pi}$.
Keeping only the leading terms, we obtain
\be
  \theta_1(w/L)=2{\rm e}^{-\pi /4}(\sin\pi w/L-{\rm e}^{-2\pi}
    \sin 3\pi w/L+\dots)
\ee
\begin{eqnarray}
    \sum_{\nu=1}^4|\theta_\nu(w/2L)|=2(1&&+{\rm e}^{-\pi/4}(\sin\pi w/2L
    +\cos\pi w/2L)\nonumber\\
    &&-{\rm e}^{-9\pi/4}(\sin 3\pi w/2L
    -\cos 3\pi w/2L)+\dots).
\end{eqnarray}
Using $w=u+iv$ and defining the symbols
\begin{eqnarray}
\sc[u,v]&&=\sin(\pi u/L)\cosh(\pi v/L)\nonumber\\
\cs[u,v]&&=\cos(\pi u/L)\sinh(\pi v/L)
\end{eqnarray}
and similar expressions for $\ss[u,v]$ and $\cc[u,v]$ which enable a 
compact notation for $\sin\pi w/L=\sc[u,v]+i\cs[u,v]$,
the expression of $\zeta(w)$ in equation~(\ref{eq-varpi}) is given in terms of trigonometric
functions:
\begin{eqnarray}
\fl|\theta_1(w/L)|=2q\left[
(\sc[u,v]-q^8\sc[3u,3v])^2 
+(\cs[u,v]-q^8\cs[3u,3v])^2
\right]^{1/2}\nonumber\\
\fl\left(\sum_{\nu=1}^4|\theta_\nu(w/2L)|\right)^2
=2\left[1+q(\sc[{\textstyle \frac u2},{\textstyle\frac v2}]
+\cc[{\textstyle \frac u2},{\textstyle\frac v2}])
-q^9(\sc[{\textstyle \frac {3u}{2}},{\textstyle\frac{3v}{2}}]
-\cc[{\textstyle\frac{3u}{2}},{\textstyle\frac{3v}{2}}])\right]^2\nonumber\\
+2\left[1+q(\cs[{\textstyle \frac u2},{\textstyle\frac v2}]
-\ss[{\textstyle \frac u2},{\textstyle\frac v2}])
-q^9(\cs[{\textstyle\frac{3u}{2}},{\textstyle\frac{3v}{2}}]
+\ss[{\textstyle\frac{3u}{2}},{\textstyle\frac{3v}{2}}])\right]^2
\end{eqnarray}
It is instructive to recover the cylinder limit from the torus geometry.
For a square of arbitrary aspect ratio $L'/L$, the expansion parameter  $q$ in the
previous series expansions has to be replaced by  ${\rm e}^{-\pi L'/L}$, and the larger the aspect ratio,
the faster the series converges. Let us consider the limit $L'\to\infty$, $L$ fixed. 
Keeping only the leading terms and omitting unimportant prefactors, we have
\be|\theta_1(w/L)|\simop_{L'/L\to\infty} |\sc^2[u,v]+\cs^2[u,v]  |^{1/2},\ee hence from
$\zeta(w)\sim |\theta_1(w/L)|$ it follows the traditional expression~\cite{Cardy84DL}
\be
\langle\bsigma_0\cdot\bsigma_w\rangle_{\pbc}
\sim \left[\cosh\left(\frac{2\pi v}{L}\right)-\cos\left(\frac{2\pi u}{L}\right)
\right]^{-\frac 12\eta_\sigma},
\ee
which leads to equation~(\ref {eq-corrstrip}) in the limit $v/L\gg 1$\ftnote{6}{In the standard notation, the roles of 
$u$ and $v$ are reversed, because we have considered here the infinite direction of the cylinder in the imaginary
direction.}.

We can then plot the correlation function data directly versus the
rescaled variable $\zeta(w)$
on a log-log scale, as shown in figure~\ref{fig-corrPbc}.
On this scale, a linear
behaviour would indicate a power-law decay.
A deviation from the straight line can  be observed at large values of $\zeta(w)$ where
the data are scattered. The quality of the data is unfortunately not improved by a better statistics (ten
times more iterations).
This correction is more pronounced as the KT point is approached where one knows 
that logarithmic corrections are present. 
From this figure, it is clearly difficult to get a reliable estimation of
the exponents, but we nevertheless performed a fit in the range $\zeta(w)<4.10^{-4}$ where the
behaviour is closer to a pure power law\ftnote{7}{This range does not correspond to the 
right limit for an investigation of the asymptotic behaviour of the correlation function
(especially when logarithmic corrections
may be involved) , since
the limit $|z|\to\infty$ in the plane corresponds to $\zeta(w)\to \infty$.}.
The straight lines reported there
correspond to (power-law fits with) slopes $\eta_\sigma(T)$
as given in  table~\ref{tab1} in the conclusion. 
They match perfectly the correlation function data in the regime of small
$\zeta(w)$, therefore if they do not confirm nicely the expression in  
(\ref{eq-corr-torus}), they at least do not contradict this equation. 
A stronger support to equation~(\ref{eq-corr-torus}) is the fact that 
the values of the exponents reported in table~\ref{tab1}
are consistent with more refined estimates
deduced from the fit of the
order parameter profile in section~\ref{sec:fitprofile} (the reader can compare columns
{\it a} and {\it b} in table~\ref{tab1} to convince himself of the compatibility of the results).

\subsubsection{Scaling of correlations in open square:}
Simulations are then performed according to a similar procedure, but now
in an open square with free BC along the four edges.
Equation~(\ref{eqkappa}) may be used to estimate the surface exponent
$\eta_\|(T)$. Plotting the quantity
\be
\langle\bsigma_{w_1}\cdot\bsigma_w\rangle_{\fbc}\times
[\kappa(w)]^{\frac 12\eta_\sigma(T)}=\psi(\omega)\ee
as a function of the variable $\omega$ defined in 
equation~(\ref{eq-omega}), and given in terms of the $w$-coordinates as
follows:
\be
\omega=\Im\left[\sn\frac{2{\rm K}w_1}{L}\right]
\times\Im\left[\sn\frac{2{\rm K}w}{L}\right]\times
\left|\sn\frac{2{\rm K}w_1}{L}-\sn\frac{2{\rm K}w}{L}
\right|^{-2}\label{eq-omegaSq}
\ee
will give access to the universal function $\psi(\omega)$.
Here, the values of $\eta_\sigma(T)$ are taken from the previous section (column {\it a} in table~\ref {tab1}).
As we already mentioned above, these are not our most precise determinations
of the bulk scaling dimensions, but the deviation with the forthcoming results of the following section
being small (see table~\ref{tab1}), 
there is no dramatic consequence to use them.

\begin{figure} [ht]
\vspace{0.2cm}
        \epsfxsize=10cm
        \begin{center}
        \mbox{\epsfbox{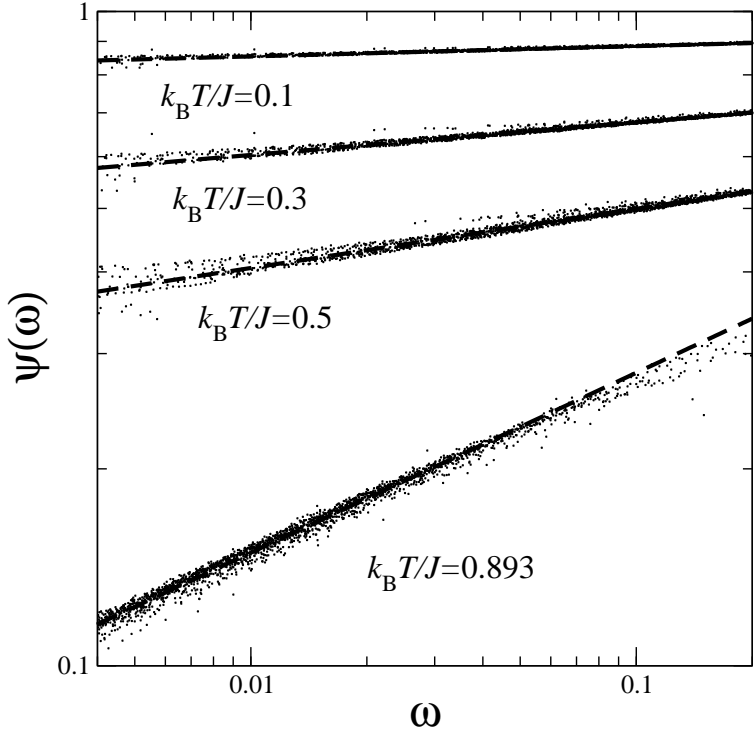}\qquad}
        \end{center}
        \caption{Log-log plot of the universal scaling function
        $\psi(\omega)$ at the KT point and below. The dashed lines represent a fit
      leading to the surface correlation function exponents ($L=101$).}
        \label{fig-psiomega}  \vskip -0cm
\end{figure}

The function $\psi(\omega)$ is expected to behave as
\be\psi(\omega)\simop_{\omega\ll 1}\omega^{\frac 12\eta_\|(T)},\ee
and leads to a determination of the surface exponent as a function of the temperature.
In the case of the $2d$ Ising model for example, the scaling function was calculated analytically
by Cardy~\cite{Cardy84},
\begin{eqnarray}
\psi_{\rm Ising}(\omega)&=&\sqrt{(1+4\omega)^{1/4}+(1+4\omega)^{-1/4}}\nonumber\\
&\sim&\omega^{1/2}(1-\omega+{\textstyle \frac 94}\omega^2+\dots),
\end{eqnarray}
in accordance with $\eta_\|^{\rm Ising}=1$.

The rescaled space variables $\kappa(w)$ and $\omega$ involve 
Jacobi elliptic sine. The series expansion is the following,
\be
\sn\frac{2{\rm K}w}{L}=\frac{2\pi}{k{\rm K}}
\sum_{n=0}^\infty \frac{{\rm e}^{-\pi(n+1/2)}}{1-{\rm e}^{-\pi(2n+1)}}
\sin\left[\frac{(2n+1)\pi w}{L}\right].
\ee
and the plot  of $\psi(\omega)$ is shown in figure~\ref{fig-psiomega} at 
several temperatures below  and at the KT transition.
The surface scaling dimension $\frac 12\eta_\|(T)$ is deduced from a power-law
fit at small values of $\omega\le 0.03$ (dashed lines)\ftnote{8}{This is the right limit this time, 
since $|z|\to\infty$ corresponds to $\omega\to 0$.}.
At the Kosterlitz-Thouless transition, we can also fit the scaling function for comparison
with the Ising case. One obtains
\be
\psi_{{\rm KT}}(\omega)\sim\omega^{0.274}(1-0.401\omega+0.221\omega^2+\dots).
\ee

\subsection{Behaviour of the density profiles\label{sec:fitprofile}}
\subsubsection{Scaling of $m_{\Fbc}(w)$:}
\begin{figure} [ht]
        \epsfysize=5.cm
        \mbox{\epsfbox{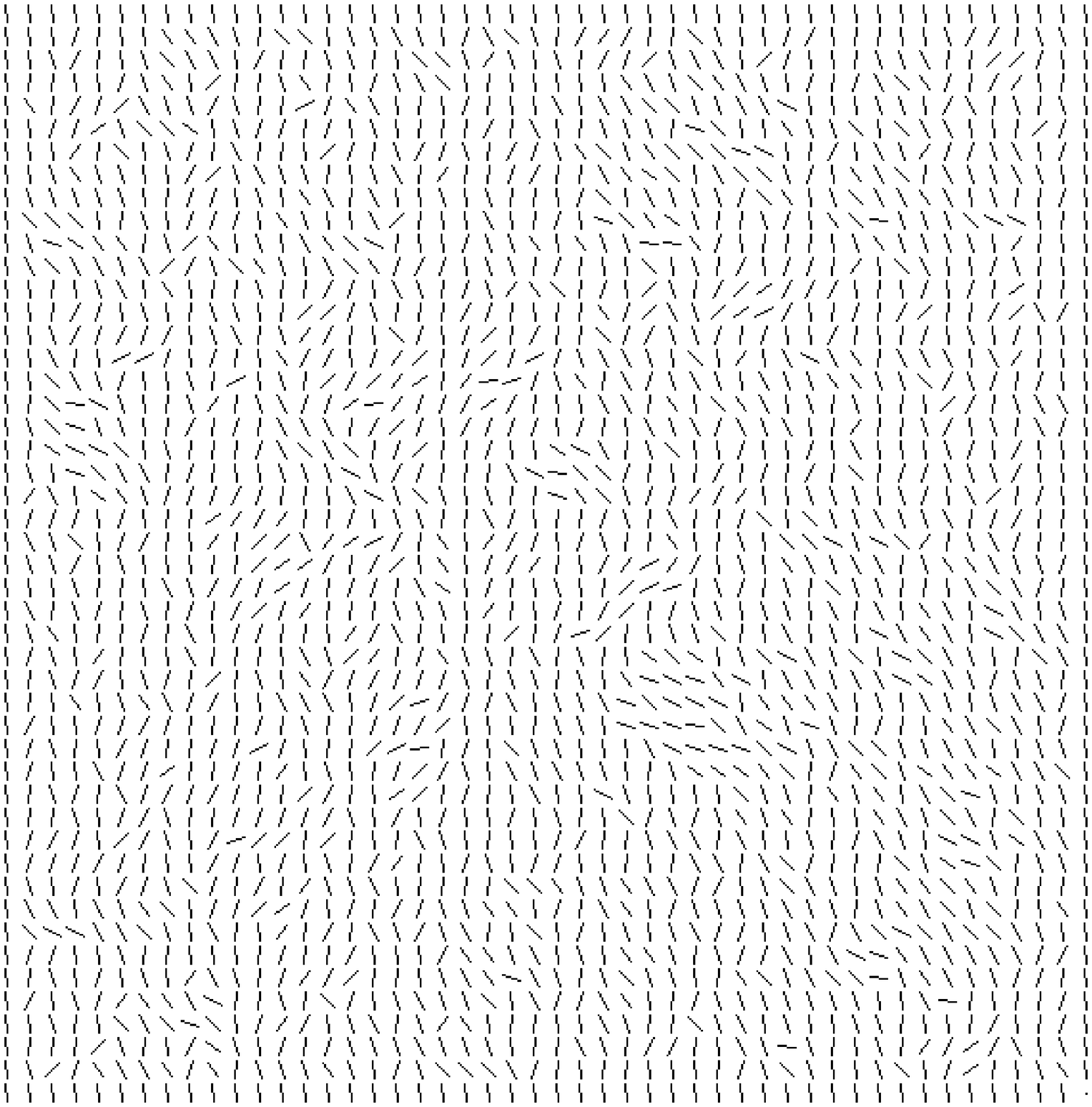}\qquad}\hfill
\vskip-4.8cm
        \epsfysize=4.3cm
\hfill\mbox{\epsfbox{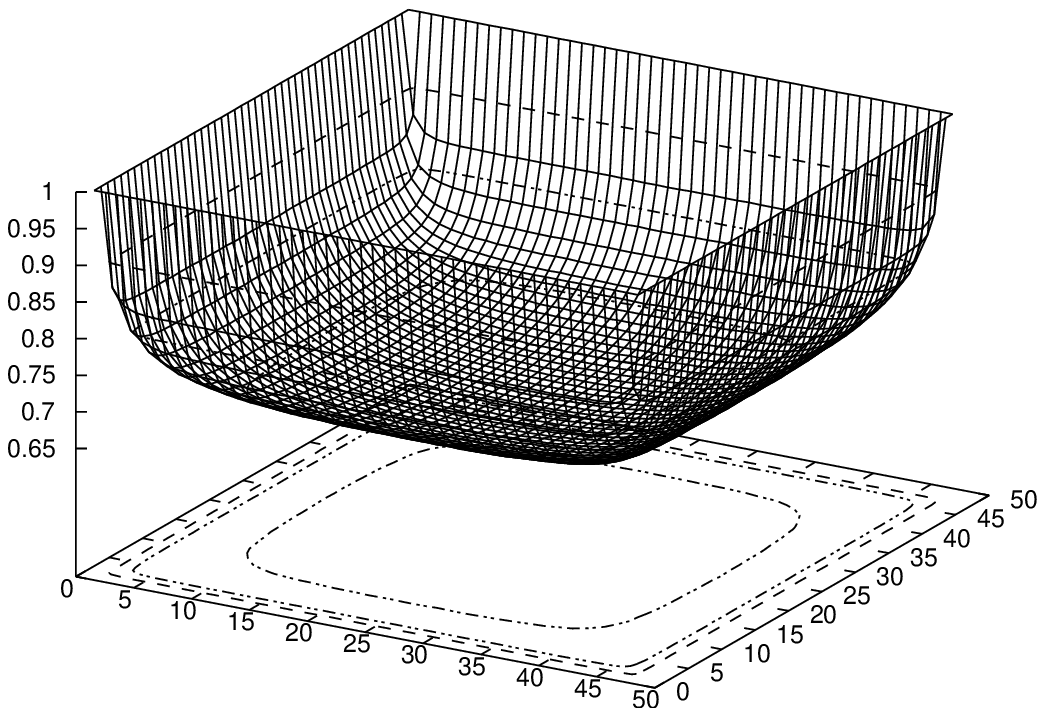}}\qquad
\vskip 1.cm
        \caption{Monte Carlo simulations of the $2d$ $XY$ model inside a square
        $48\times 48$ spins  with fixed boundary conditions 
        below the Kosterlitz-Thouless
        transition temperature. Typical configuration on the left (where the fixed BC
        are easily observed by the common direction taken by all the border spins)
        and average over $10^6$ MCS/spin after cancellation of $10^6$ for
        thermalization (cluster update algorithm) on the right.}
        \label{fig:0}  \vskip 0.2cm
\end{figure}

Density profiles present the advantage of being one-point functions.
They are thus supposed to be determined more precisely numerically than
two-point correlation functions and should therefore lead to more refined
estimations of the critical exponents. 
We start with simulations inside an open square with fixed boundary 
conditions ($\Fbc$), according to the prescriptions given in the description of the
algorithm.  
The order parameter profile $m_{\Fbc}(w)$ is defined according 
to the definition of a reference orientation on the fixed boundaries, i.e.
\begin{eqnarray}
m_{\Fbc}(w)&=&\langle\bsigma_w\cdot\bsigma_{\partial\Lambda (w)}
\rangle_{\Fbc}\nonumber\\
&=&\langle\cos(\theta_w-\theta_{\partial \Lambda(w)})\rangle_{\Fbc}.
\end{eqnarray}
After averaging over the `production sweeps', 
one gets a characteristic smooth profile as shown in 
figure~\ref{fig:0} which also presents a typical configuration with fixed BC.

\vspace{-0.0cm}

\begin{figure} [ht]
\vspace{0.2cm}
        \epsfxsize=10cm
        \begin{center}
        \mbox{\epsfbox{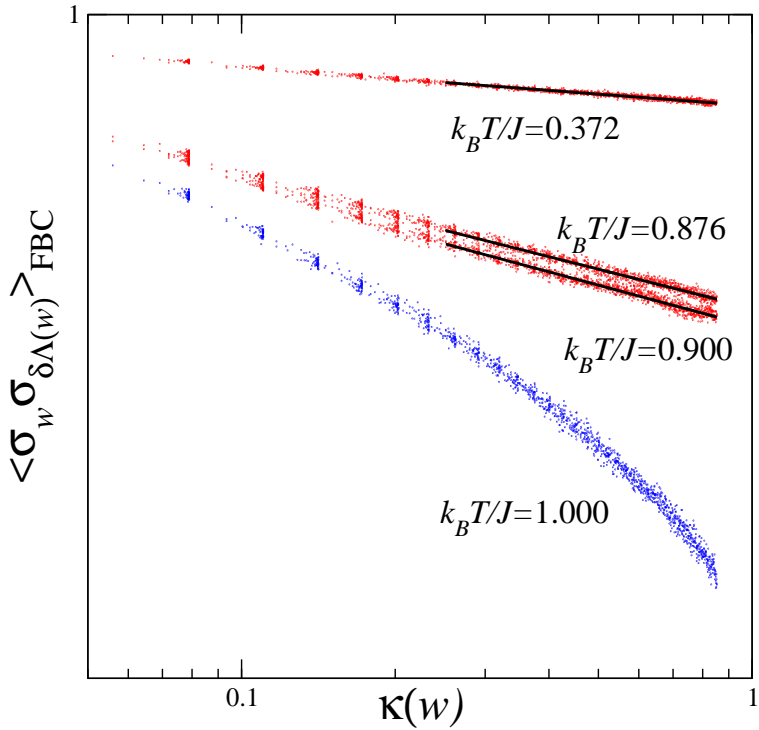}\qquad}
        \end{center}
        \caption{Monte Carlo simulations of the $2d$ $XY$ model inside a square
        $100\times 100$ spins ($10^6$ MCS/spin after cancellation of $10^6$ for
        thermalization, cluster update algorithm). The 
        figure shows the rescaled order parameter density against the rescaled
        distance on a log-log scale at  different 
        temperatures below the Kosterlitz-Thouless
        transition temperature and slightly above. 
        }
        \label{fig:1}  \vskip -0cm
\end{figure}

A log-log plot of $m_{\Fbc}(w)$ with respect to the reduced 
variable $\kappa(w)$ is shown in figure~\ref{fig:1}
at two different temperatures below $T_{\rm KT}$, 
roughly at $T_{\rm KT}$, and one temperature
slightly above. 
A first observation is the confirmation of the functional form of
equation~(\ref{eqParamOrder}) in the whole low-temperature phase. One indeed observes 
a very good data collapse of the $L^2$ points
onto a single power-law  master curve (a straight line on this scale). 
The next information is the rough confirmation of the value
of the critical temperature, since above $T_{\rm KT}$, 
the master curve is no longer a straight line,
indicating that the corresponding decay in the half-infinite geometry differs 
from a power-law as it should in the high-temperature 
phase\ftnote{9}{One should nevertheless mention that this does not lead 
to a precise determination of $T_{\rm KT}$,
since at $k_BT/J=0.9$ for example, the master curve is hardly distinguishable 
from a straight line.}. 
In reference~\cite{BercheFarinasParedes02}, we have shown how the
computation the $\chi^2$ per d.o.f. as a function of temperature
is an indicator of the location of the KT transition. The $\chi^2$ has a very
small value, revealing the high quality of the fit, 
in the low-temperature phase and then increases 
significantly above $T_{\rm KT}$ when the behaviour of the 
density profile is no longer algebraic.

We have also shown that there is probably
no logarithmic correction at the Kosterlitz-Thouless point, since the curve of
figure~\ref{fig:1} does not seem to differ significantly from a true
power-law. Although not a proof, a numerical evidence of that is the plot of
$m_{\Fbc}(w)\times[\kappa(w)]^{1/8}$ {\it vs} $\kappa(w)$ which remains constant 
on the whole range of values of $\kappa(w)$~\cite{BercheFarinasParedes02}.
This observation makes the order 
parameter profile with fixed boundary conditions a very convenient quantity
for the determination of the bulk correlation functions exponent, since it
displays a pure algebraic decay up to the Kosterlitz-Thouless transition 
point\ftnote{1}{The decay exponent,
if not fixed, but determined numerically, leads to a quite good result 
(for example for a size $L=100$, we get 
$\eta_\sigma(T_{\rm KT})=0.250(28)$).}.

Fitting the curves of figure~\ref{fig:1} to power-laws 
leads to the values of the scaling
dimension $\eta_\sigma(T)$ in the whole critical region $T\le T_{\rm KT}$
as shown in figure~\ref{fig:ccl} in the conclusion. The data are also reported in table~\ref{tab1} in
the conclusion. The accuracy is much better than in the case of the two-point correlation
functions.

\subsubsection{Scaling of $m_{\Ffbc}(w)$:}
\begin{figure} [th]
  \ \vskip2cm
        \epsfysize=5.5cm
        \mbox{\epsfbox{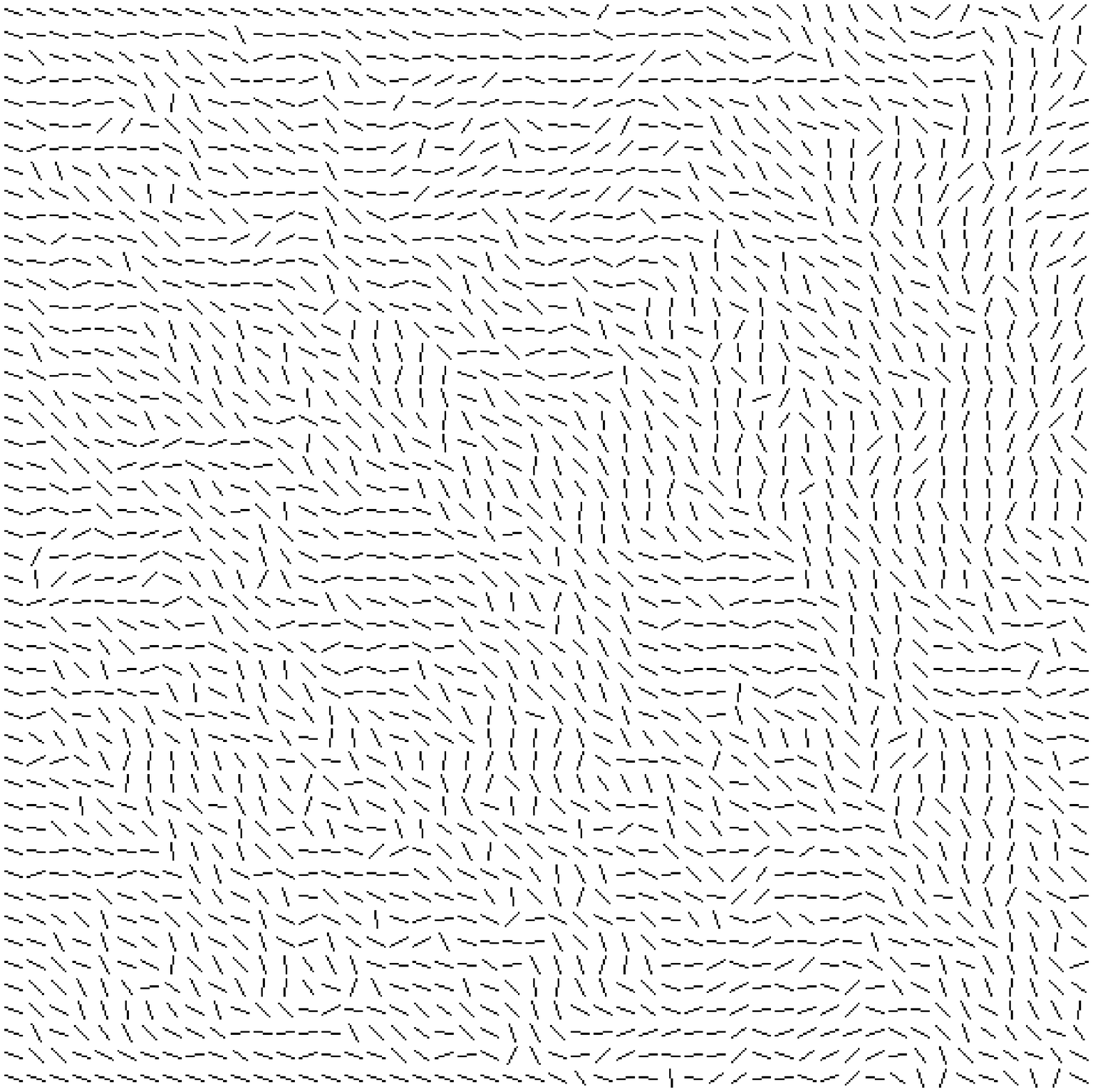}\qquad}\hfill
        \vskip-7cm
        \epsfysize=4cm
        \hfill\mbox{\epsfbox{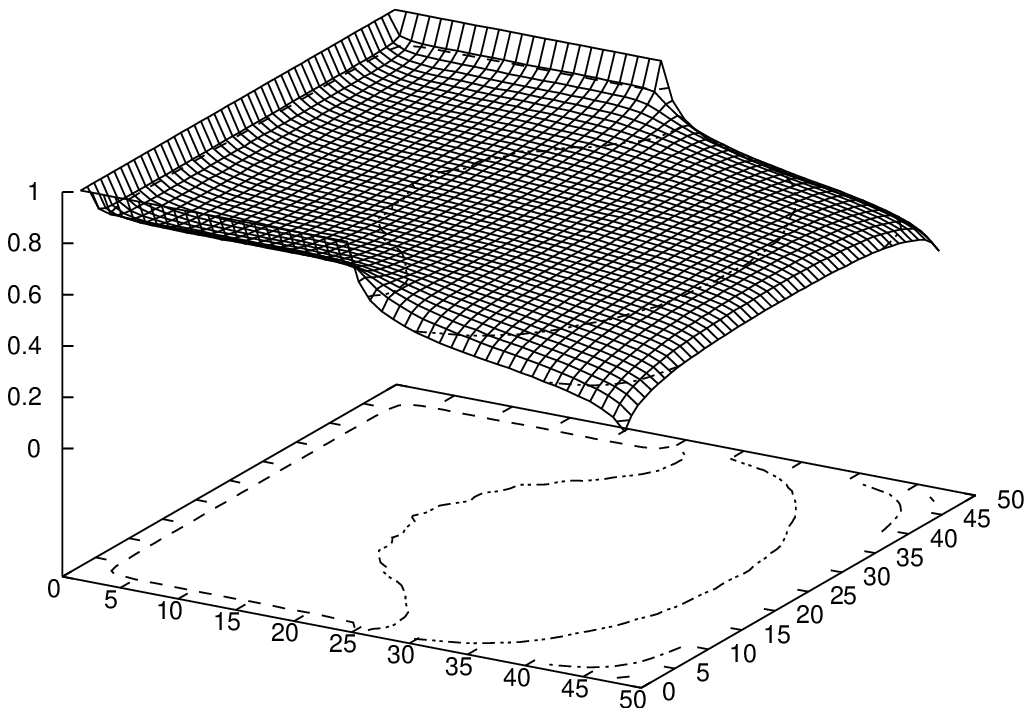}} 
        \vskip1mm
        \epsfysize=4cm
        \hfill\mbox{\epsfbox{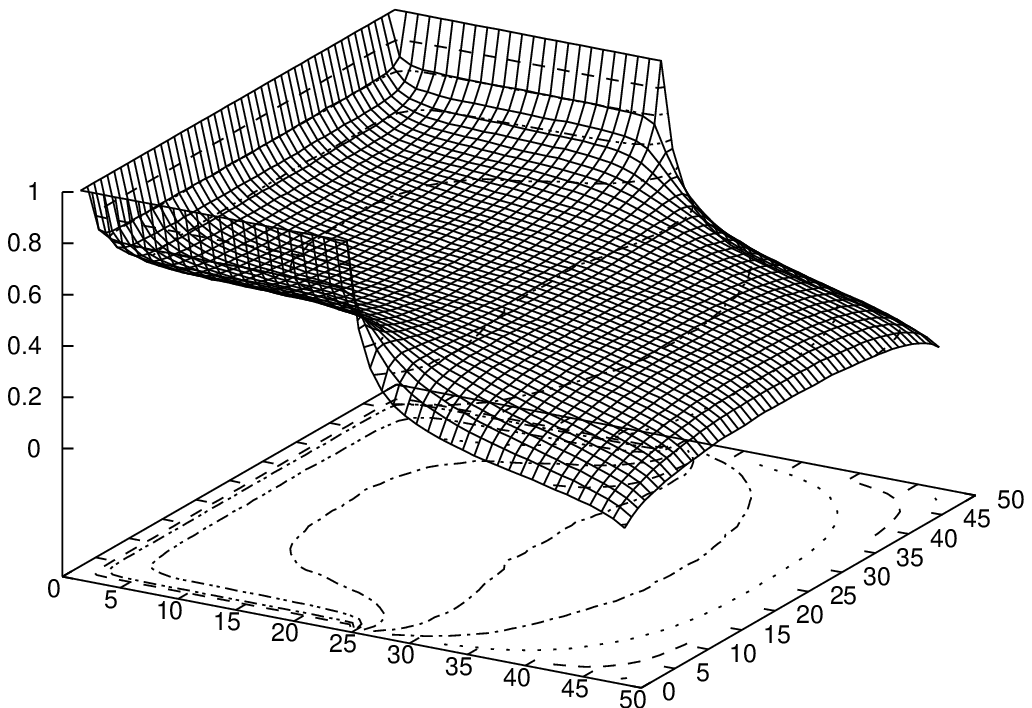}}
        \vskip 0.5cm
        \caption{Monte Carlo simulations of the $2d$ $XY$ model inside a square
        $48\times 48$ spins  with mixed 
        fixed-free boundary conditions below the Kosterlitz-Thouless
        transition temperature. Typical configuration on the left
        and average over $10^6$ MCS/spin after cancellation of $10^6$ for
        thermalization (cluster update algorithm) 
        at $k_BT/J=0.488$ and $k_BT/J=0.915$ on the right.}
        \label{fig:2}  \vskip 0.2cm
\end{figure}

For this last choice of boundary conditions, hereafter referred to as 
``fixed-free BC'' (and denoted by $\Ffbc$), the open square has
half of its boundaries which are left free while the other half is kept fixed.
This is illustrated in figure~\ref{fig:2} where the averaged profile is also shown at two different
temperatures. The shape of the plotted surface helps understanding the BC.

\vspace{-0.0cm}

In the previous section, we reported some evidence that the 
logarithmic corrections at 
the KT point are negligible for the order parameter profile with fixed BC. 
Assuming that there
will be no substantial difference for fixed-free BC, we perform a fit 
of $\ln m_{\Ffbc}(w)$ against the 
two-dimensional linear expression, 
${\rm const}- \frac 12{\eta_\sigma}(T)\ln \kappa(w)+\frac 12{\eta_\|}(T)\ln\mu(w)$, 
as expected from equation~(\ref{eqkappamu}). 
The value obtained for the bulk correlation function 
exponent $\eta_\sigma(T)$ (already known precisely from previous section) 
will be a test for the reliability of the fit.
The simulations are performed at several temperatures, up to 
$k_BT_{\rm KT}/J\simeq 0.893$.
The three parameters linear fits lead to correct determinations
of both exponents (we studied  different system sizes 
ranging between $L=48$ and $L=100$ and  even $200$ at the KT point). 
The value of the bulk exponent is in a good agreement with the determination
which follows from the fit of the profile with fixed BC, and the surface exponent 
also fits nicely to the values deduced from the two-point correlation function with free
BC.

\subsubsection{Scaling of $\Delta\varepsilon_{\Fbc-\fbc}(w)$:}
We concentrated mainly up to now on the magnetic properties. The thermal exponents are
also interesting quantities which characterise a critical point. In the case 
of the $XY$ model, the temperature is a marginal field, responsible for the
existence of a critical line in the whole low-temperature phase. It thus
implies a thermal scaling exponent $x_\varepsilon=d-y_t=2$ which ensures
a vanishing RG eigenvalue $y_t=0$ (up to $T_{\rm KT}$ where it is consistent 
with the essential singularity of $\xi$
above the KT point). The energy-energy correlation function
should thus decay algebraically as
\be
\langle\varepsilon_{z_1}\varepsilon_{z_2}\rangle\sim
|z_1-z_2|^{-\eta_\varepsilon},\ee
with $\eta_\varepsilon(T)=2x_\varepsilon=4$ $\forall T<T_{\rm KT}$.

We check qualitatively this expression of the energy-energy correlations from the behaviour of
the energy density profile. 
The energy density at site $w$ is for example defined as the average value of the energies
of the four links reaching $w$:
\be
\varepsilon_w=\frac {1}{2d}\sum_{\hat\mu}(\bsigma_{w-\hat\mu}\cdot\bsigma_{w}+
\bsigma_w\cdot\bsigma_{w+\hat\mu}).
\ee

\begin{figure} [ht]
\vspace{0.2cm}
        \epsfxsize=10.5cm
        \begin{center}
        \mbox{\epsfbox{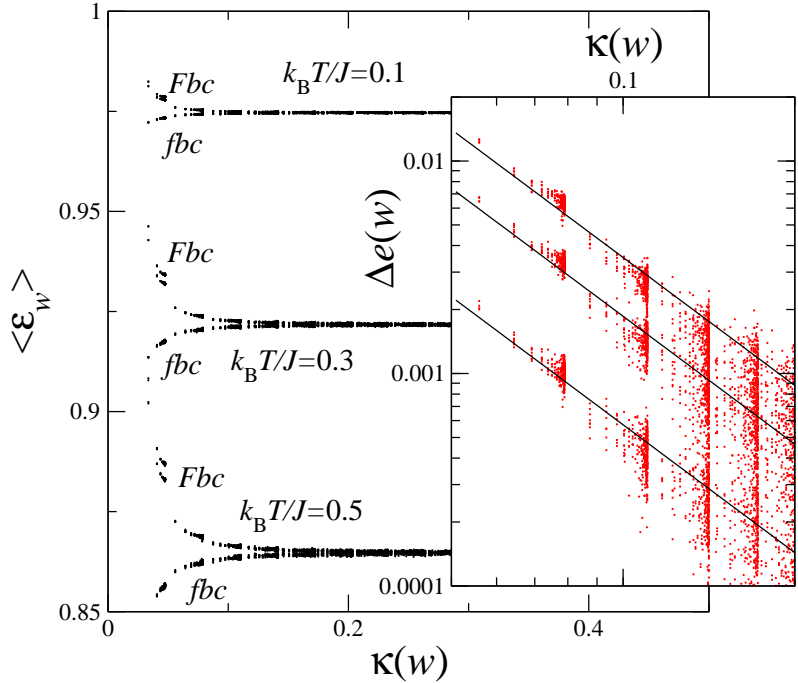}\qquad}
        \end{center}
        \caption{Monte Carlo simulations of the $2d$ $XY$ model inside a square
        $100\times 100$ spins ($10^6$ MCS/spin after cancellation of $10^7$ for
        thermalization, cluster update algorithm). The main 
        figure shows the local energy density {\it vs} the rescaled variable $\kappa(w)$
        at different temperatures and for different BC. The insert is a log-log plot of the 
        difference $\Delta e(w)$ as explained in the text.
        }
        \label{fig_Delta_energy}  \vskip -0cm
\end{figure}

In comparison with the magnetisation profile,
a difficulty occurs due to the existence of a regular contribution in the 
energy density. This term cancels after a suitable difference between
profiles obtained with different BC.
For that reason, we made two different series of simulations with free 
($\fbc$) and fixed ($\Fbc$) boundary 
conditions. 
From the latter simulations we had already extracted the singular behaviour of the magnetisation 
profile  since there is no regular contribution in the scale-invariant phase,
\begin{equation}
m_{\Fbc}(z)={\cal A}_m(T) y^{-\eta_\sigma(T)/2},
\label{eq:M}
\end{equation}
where ${\cal A}_m(T)$ is a non-universal ``critical'' amplitude\ftnote{2}{Here we use
the term ``critical'', since the whole  low temperature phase is considered as critical.} 
which depends on the temperature. 

On the other hand, 
the two series of simulations are necessary in order to extract the singularity
associated to the energy
density which contains a non universal   
regular contribution $\langle\varepsilon_0(T)\rangle$ which depends on $T$
and a singular contribution:
\begin{eqnarray}
\langle\varepsilon_z\rangle_{\Fbc}&=&
\langle\varepsilon_0(T)\rangle+{\cal B}_{\Fbc}(T) y^{-\eta_\varepsilon(T)/2},\\
\langle\varepsilon_z\rangle_{\fbc}&=&
\langle\varepsilon_0(T)\rangle+{\cal B}_{\fbc}(T) y^{-\eta_\varepsilon(T)/2}.
\label{eq:E}
\end{eqnarray}
This is clearly illustrated in figure~\ref{fig_Delta_energy} where convergence towards the
same temperature-dependent constant $\langle\varepsilon_0(T)\rangle$ is shown.
We also observe that the amplitudes of the singular terms have opposite signs. Therefore 
a simple difference of the quantities measured in the square geometry,
\be
\Delta e(w)=\langle\varepsilon_w\rangle_{\Fbc}-\langle\varepsilon_w\rangle_{\fbc}
\sim \Delta{\cal B} \times [\kappa(w)]^{-\frac 12\eta_\varepsilon(T)}
\ee
leads to the value of the thermal scaling dimension $\eta_\varepsilon(T)$.

The insert in figure~\ref{fig_Delta_energy} shows a log-log plot of the 
difference $\Delta e(w)$ {\it vs} $\kappa(w)$ at three temperatures below $T_{\rm KT}$. 
Due to strong fluctuations,
the data scatter around straight lines which represent the theoretical slopes
$[\kappa(w)]^{-2}$. This figure, though far from being definitely conclusive,
confirms that the  exponent
of the decay of energy-energy correlations keeps a constant value
$\eta_\varepsilon(T)=4$ in the low-temperature phase of the $XY$ model.
At the KT transition, this value was confirmed very accurately  by Bl\"ote and 
 Nienhuis~\cite{BloteNienhuis89}.

\section{Discussion}
The different determinations of the bulk and surface correlation function scaling dimensions
reported in this paper are mutually consistent.
These exponents are obtained using completely independent
techniques, from four independent series of simulations (the boundary conditions, but also the
quantities which are computed are different). The fits themselves are different, since these
quantities obey specific functional expressions.

\begin{figure} [ht]
\vspace{0.2cm}
        \epsfxsize=12cm
        \begin{center}
        \mbox{\epsfbox{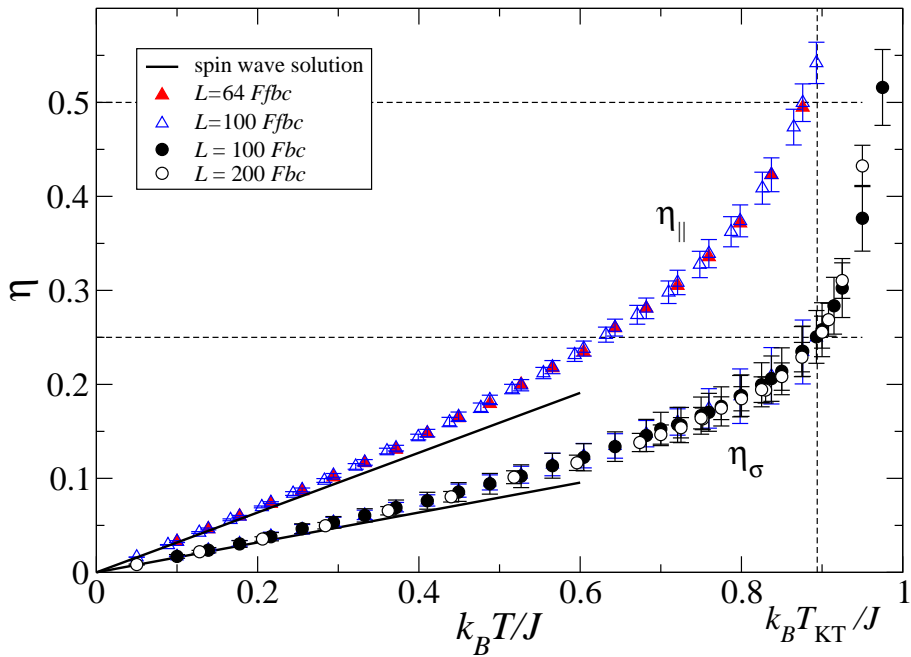}\qquad}
        \end{center}
        \caption{Bulk and surface correlation decay exponents of the $2d$ $XY$ 
          model as a function of the temperature. The values follow from independent
          linear fits of the 
          magnetisation profiles with $\Fbc$ (circles) or $\Ffbc$ (triangles).}
        \label{fig:ccl}  \vskip -0cm
\end{figure}

The exponents
are plotted in figure~\ref{fig:ccl} for different system sizes. 
Our results (for a few values of the temperature) are also given in table~\ref{tab1}. In column
{\it a}, {\it b}, and {\it c} we present the bulk exponent. Columns {\it d} and {\it e} give the surface exponent.
The error bars (for the values resulting from the fits of
the magnetisation profiles, columns {\it b}, {\it c}, {\it d}) 
correspond to one standard deviation. We have to mention here that 
these error bars should be considered with some care for at least two reasons. A  first reason is that 
we did not make any 
extrapolation to the thermodynamic limit, because the results at a few sizes
seem already very stable as it can
be observed in figure ~\ref{fig:ccl}. Although we did not analyse systematically the size effects, we believe that this is
not the most important contribution to the error.
A second reason has more drastic consequences.
This is the influence of the fitting window. 
When the range of fit is varied
in a {\it reasonable} scale in the `large $\kappa$' regime, 
it usually produces a small deviation in the resulting exponent. 
For example, fitting a power law $\sim [\kappa(w)]^x$ in the
window $0.10<\kappa(w)<1$ or $0.25<\kappa(w)<1$, lead to  compatible results within the error 
bars (in this example at the KT transition, we get respectively $\eta_\sigma=0.260(18)$
and $0.250(28)$).
If the window range is changed significantly on the other hand, we do no longer control the stability of
the results, since the errors increase considerably (for example $\eta_\sigma=0.247(92)$ for 
$0.50<\kappa(w)<1$). Because of these constraints, we believe that the values of the exponents that we measure are
reliable, but the errors are only rough estimates.

\begin{table}
        \caption{Values of the exponents of the bulk and surface correlation
          function in the spin-wave phase of the $2d$ $XY$ model. They are
        obtained after fitting the density profiles or correlation functions
      to the conformal functional expressions for a system of size $L=100$.
    We have checked that the results are already very stable at such a size.
  We mention that determinations {\it a}, {\it b} and {\it c} are completely independent for the
bulk exponent, and that determinations {\it d} and {\it e} for the surface exponent are also independent.
For determination {\it e}, the value of the bulk exponent in {\it a} is used as input. The most reliable results
are printed in bold font.}
        \label{tab1}
        \begin{indented}
        \item[]
        \begin{tabular}{llllllll}
        \br
        &\centre{7}{Correlation function exponents}  \\
        $T$ & \centre{4}{$\eta_\sigma$} & \centre{3}{$\eta_\|$} \\
        & \crule{4} & \crule{3}  \\
        & ${\it a}$ & ${\it b}$ & ${\it c}$ & S.W. & ${\it d}$ & ${\it e}$ & S.W. \\
        \mr
        0.1    & 0.017 &\bf 0.017(2)  &\bf  0.017(1) &0.016 &\bf  0.032(1)   &  0.031 & 0.032\\
        0.2    & 0.037 &\bf 0.035(4)  &\bf  0.036(3) &0.032 &\bf  0.069(1)   &  0.064 & 0.064\\
        0.3    & 0.057 &\bf 0.052(7)  &\bf  0.052(5) &0.048 &\bf  0.103(2)   &  0.100 & 0.095\\
        0.4    & 0.077 &\bf 0.073(9)  &\bf  0.074(6) &0.064 &\bf  0.143(4)   &  0.140 & 0.127\\
        0.5    & 0.103 &\bf 0.096(11) &\bf  0.100(8) &-- &\bf  0.186(5)   &  0.181 &-- \\
        0.6    & 0.127 &\bf 0.122(14) &\bf  0.122(13)&-- &\bf  0.231(8)   &  0.228 &-- \\
        0.7    & 0.154 &\bf 0.153(18) &\bf  0.155(14)&-- &\bf  0.298(12)  &  0.294 &-- \\
        0.8    & 0.194 &\bf 0.188(22) &\bf  0.187(29)&-- &\bf  0.374(17)  &  0.366 &-- \\
        0.893& 0.249 &\bf 0.250(28) &\bf  0.250(34)&-- &\bf  0.542(22)  &  0.548 &-- \\
        \br
        \end{tabular}
        \item[] $^{\it a}$ deduced from the fits of $\langle\bsigma_{0}
          \cdot\bsigma_w\rangle_{\pbc}$ ($L=100$), section 3.2.1.   
        \item[] $^{\it b}$ deduced from the fits of $m_{\Fbc}(w)$, section 3.3.1.   
        \item[] $^{\it c}$ deduced from the fits of $m_{\Ffbc}(w)$ ($L=100$ or $200$ at the KT point),
          section 3.3.2.
        \item[] $^{\it d}$ deduced from the fits of $m_{\Ffbc}(w)$ ($L=100$ or $200$ at the KT point),
          section 3.3.2.
        \item[] $^{\it e}$ deduced from the fits of $\langle\bsigma_{w_1}
          \cdot\bsigma_w\rangle_{\fbc}$ ($L=101$), section 3.2.2.   
        \item[] S.W.  gives the result of the spin wave contribution.   
        \end{indented}
\end{table}

The weaker result of the paper probably comes from the
correlation function with periodic boundary conditions which was fitted to an approximate 
functional expression. There are also in this case logarithmic
corrections and an unavoidable scattering of the data 
which prevent a fit in the interesting asymptotic region. The consequence is a bulk
exponent which is {\it a priori} obtained with a poor level of confidence, and
for this reason we decided to avoid any
error bar in column {\it a} of table~\ref{tab1}.
It is amazing to observe that in spite
of these restrictions, these results do indeed agree with the most refined values deduced from
the magnetisation density profile with fixed boundary conditions (see column {\it b} in table~\ref{tab1}).
Another consequence of our approach is again the {\it a priori}
weak level of confidence of the surface exponent 
deduced from the correlations in the open system (column {\it e} in the table), since we use the
bulk values from column {\it a} as input parameters for this latter fit.
The surface exponent in fact fits correctly the most refined values deduced from
the magnetisation density profile with mixed fixed-free boundary conditions (column {\it d}).
Obviously, the stronger results come from the magnetisation profiles analyses.  We believe also that another result
of interest is the confirmation that the functional expressions derived in section 2 are valid and are in fact
very efficient for the numerical investigation of critical properties of two-dimensional systems. Such
applications of conformal mappings are of course well known in the case of strip geometries with e.g. the
celebrated correlation length amplitude-exponent relation\cite{Cardy84DL}.

Let us now discuss briefly the results.
In the low-temperature limit, one knows that the spin-wave description captures the essential of
the physics of the problem. From this limit, the linear behaviour of the bulk exponent with the
temperature, $\eta_\sigma (T)=k_BT/2\pi J$, is easily recovered.
It is of course reasonable to expect that the low-temperature limit of the surface exponent can also
be recovered by the spin wave approximation (see appendix). Indeed, the 
surface properties of the Gaussian model
at the ordinary transition were studied by Cardy~\cite{Cardy84}. They describe to the low-temperature
regime of the $XY$ model. 
The surface-bulk correlation function exhibits an algebraic decay 
\be
\langle\cos(\theta_0-\theta_{\bf r})\rangle_{\uhp}\simeq
\left(\frac{\pi r}{a}\right)^{-1/\pi K}
\ee
and the corresponding surface exponent is thus given in the spin wave approximation
by the linear behaviour
\be
\eta_\|(T)=\frac{k_BT}{\pi J},\quad T\to 0.
\ee
This result is simply twice the bulk value, $\eta_\|(T)=2\eta_\sigma(T)$.
This relation fits the numerical data at low temperature as shown in figure~\ref{fig:ccl} or in the table.

At the KT transition,
the bulk exponent is fully coherent with the result of Kosterlitz and Thouless
RG analysis $\eta_\sigma(T_{\rm KT})=1/4$. 
For the surface, the value 
$\eta_\|(T_{\rm KT})\simeq 0.54$ is close to the known result $\frac 12$~\cite{Cardy84}, 
a value which cannot 
be excluded from our analysis (though
out of the error bar), since the uncertainties that we quote are only indications.
It is also coherent with the surface exponent
obtained at the KT point 
of the quantum Ashkin-Teller chain ($\epsilon=-2^{-1/2}$ in reference~\cite{CarlonLajkoIgloi01}),
$\eta_\|(T_{\rm KT})=2\pi^{-1}{\rm arccos}\ \!(-\epsilon)=1/2$.
This value is furthermore in agreement with the relation 
$\eta_\|(T)=2\eta_\sigma(T)$, provided that it is valid up to the KT point.
This relation was established in the spin-wave approach, and to
extend it to the vortex dominated regime, closer to $T_{\rm KT}$, 
we note that the effect of vortices (when $T\to T_{\rm KT}^-$) when treated in the
Coulomb gas approach~\cite{Villain75,ItzyksonDrouffe89} 
is only to increase the effective temperature in the spin wave approximation, hence to enhance the
decay of the correlations. 
Like $\eta_\|(T)=2\eta_\sigma(T)={k_BT}/{\pi J}$ is correct in the spin wave treatment,
the relation $\eta_\|(T_{\rm eff})=2\eta_\sigma(T_{\rm eff})={k_BT_{\rm eff}}/{\pi J}$ should
hold in the whole critical phase, with a shift in temperature which depends on the vortices interactions,
\be
k_BT_{\rm eff}=k_BT-\frac 12\pi^2\sum_{|{\bf r}|>a}r^2\langle q(0)q({\bf r})\rangle,\ee
where $\langle q(0)q({\bf r})\rangle$ are the vortex intensity correlations (which are negative in the
neutral low temperature phase).
It is thus reasonable to expect that the relation $\eta_\|(T)=2\eta_\sigma(T)$ is exact for the whole
critical phase of the model. It can be checked that the numerical values in table~\ref{tab1} 
support this expression.

\begin{figure} [t]
\vspace{0.2cm}
        \epsfxsize=12cm
        \begin{center}
        \mbox{\epsfbox{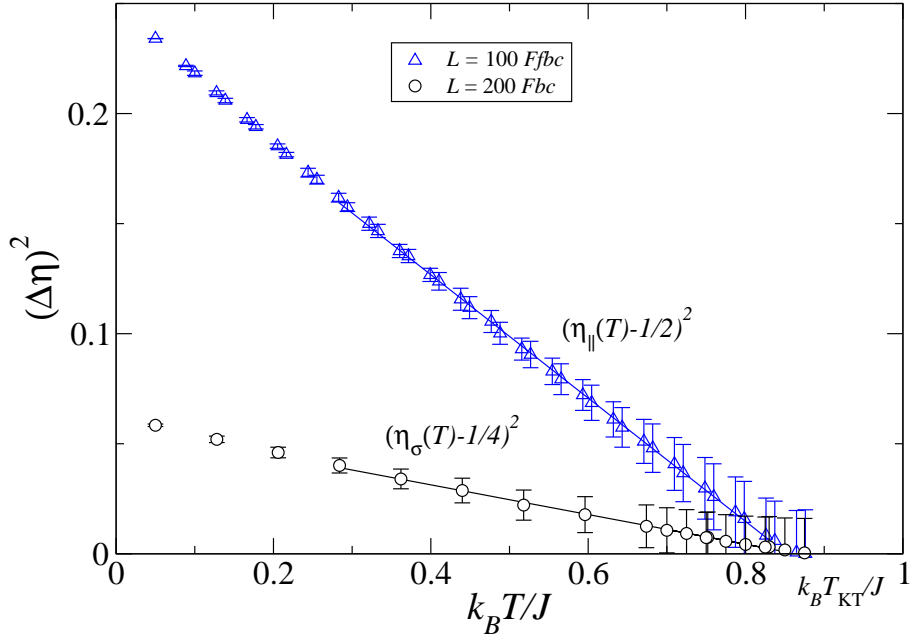}\qquad}
        \end{center}
        \caption{Square root behaviour of the bulk and surface correlation function exponents as the
        Kosterlitz-Thouless point is approached.}
        \label{fig:ccl2}  \vskip -0cm
\end{figure}

The approach of the KT point, where the temperature dependence of the 
bulk correlation function exponent is known,   is also interesting.
From the numerical data, $\eta_\sigma(T)$ can be expanded close to the 
KT point $T\to T_{\rm KT}^-$, showing a leading square root 
behaviour~\cite{KosterlitzThouless78,GuptaBaillie92} as illustrated in figure~\ref{fig:ccl2}.
The same behaviour is obtained for the surface exponent.
One should nevertheless mention here that the plot of $(\eta_\sigma(T)-\frac 14)^2$ or
$(\eta_\|(T)-\frac 12)^2$ against $T$ seem to indicate a slightly lower value of $T_{\rm KT}$ than
0.893. A clarification of that point is beyond the scope of this paper, since it would definitely require simulations
of really large systems in order to rule out the possibility of finite-size dependence of the measured
quantities.

Eventually, we mention that our analysis of the energy density profiles is compatible with the marginality
condition of the temperature in the low temperature phase of the $XY$ model. This condition, equivalent
to the continuous variation of the magnetic scaling dimension, indeed implies that
the thermal scaling dimension keeps a constant value $x_\varepsilon=2$ which do not depend on the
temperature, and this value enable to fit reasonably the numerical data.

In this paper, we performed an accurate study of the low-temperature phase of the $2d$ $XY$
model using  efficient techniques, based on a conformal mappings of the correlation
functions and density profiles. 
At the Kosterlitz-Thouless transition, the results are in agreement with the  
general picture of the RG analysis and the critical
exponents describing the algebaric decay of the correlations in the low temperature phase are
measured.

\ack 
This work has been supported by the  CINES Montpellier under project
\# c20020622309.
I would like to thank Dragi Karevski and Christophe Chatelain
for stimulating discussions and coffee breaks 
and Ferenc Igl\'oi and Ricardo Paredes for useful correspondence.

\vskip4mm
\noindent{\bf Note added in proof}
\vskip4mm
\noindent I gratefully acknowledge L.N. Shchur who drew my attention on a recent paper where similar
results were reported~\cite{ResStraley00}.

\appendix
\section*{Appendix}
In this appendix, we recall the surface-bulk correlation function obtained in the spin wave approximation.
The correlation function between a spin close to the surface
($\bsigma_0$) and another spin far in the bulk ($\bsigma_z$) is defined according to
$\langle\bsigma_0\cdot\bsigma_z\rangle_{\uhp}=\langle\cos(\theta_0-\theta_z)\rangle_{\uhp}$,
where the subscript $\uhp$ means here that $\Im z>0$. 
We use a standard notation ${\bf r}$ for $z$ from now on. 
As it is well known, in the harmonic approximation, the Hamiltonian becomes quadratic and the thermal
average leads to the Gaussian
model which implies that
\be
\langle\cos(\theta_0-\theta_{\bf r})\rangle_{\uhp}={\rm e}^{
 -{\textstyle \frac 12}\langle(\theta_0-\theta_{\bf r})^2\rangle_{\uhp}}.
\ee 
In Fourier space, $\theta_{\bf r}=N^{-1/2}\sum_{\bf q}\exp(-i{\bf qr})\theta_{\bf q}$, ($N=L^2$), 
the Hamiltonian becomes (the spins are located on the vertices of a square lattice of spacing unit 
$a$)\ftnote{3}{In contrast with the standard calculation in the whole plane, an extra factor
  $\frac 12$ appears from the orthogonality relation in the \uhp.}
\begin{eqnarray}
-\frac{H}{k_BT}&\simeq& {\rm const}-\frac 14 K\sum_{\bf q}2(2-\cos aq_x-\cos aq_y)
|\theta_{\bf q}|^2\nonumber\\
&\simeq& {\rm const}-\frac 14 K\sum_{\bf q} a^2|{\bf q}|^2
|\theta_{\bf q}|^2
\end{eqnarray}
an leads to a partition function
\be
Z_\beta=\prod_{{\bf q}}\left(\frac{4\pi}{Ka^2|{\bf q}|^2}\right)^{1/2}.
\ee
Hence, the quadratic fluctuations of the Fourier amplitudes are linear in $T$,
\be
\langle|\theta_{\bf q}|^2\rangle=\frac{2}{Ka^2|{\bf q}|^2},
\ee
and the integral ($N^{-1}\sum_{\bf q}\to\frac{a^2}{4\pi^2}\int{\rm d}{\bf q}$)
\be
\langle(\theta_0-\theta_{\bf r})^2\rangle=
\frac{a^2}{2\pi^2}\int{\rm d}{\bf q}
(1-\cos{\bf qr})\langle|\theta_{\bf q}|^2\rangle,
\ee
is easily evaluated, leading to
\begin{eqnarray}
\langle(\theta_0-\theta_{\bf r})^2\rangle&=&
\frac{1}{\pi^2K}\int_0^{\pi/a}\frac{{\rm d}{q}}{{q}}\int_0^{2\pi}{\rm d}\varphi
(1-\cos{qr\cos\varphi})\nonumber\\
&=&\frac{1}{\pi K}\int_0^{\pi r/a}\frac{1-J_0(u)}{u}{\rm d}u\nonumber\\
&\simeq&\frac{1}{\pi K}\ln \frac{\pi r}{a},
\end{eqnarray}
where a cutoff $\sim\pi/a$ was introduced.
The surface-bulk correlation function follows:
\be
\langle\cos(\theta_0-\theta_{\bf r})\rangle_{\uhp}\simeq
\left(\frac{\pi r}{a}\right)^{-1/\pi K}
\ee
and the corresponding surface exponent is thus given in the spin wave approximation
by the linear behaviour
\be
\eta_\|(T)=\frac{k_BT}{\pi J},\quad T\to 0.
\ee


\end{document}